\documentclass[useAMS,usenatbib]{mn2e}
\usepackage{color}
\usepackage{array} 
\usepackage{amsmath}
\usepackage{booktabs}
\usepackage{graphicx}
\usepackage{epstopdf}
\usepackage[labelfont=bf,labelsep=period,justification=raggedright,singlelinecheck=false,font=small]{caption}
\usepackage{wrapfig}
\usepackage{float}
\usepackage{amssymb}
\usepackage[normalem]{ulem}
\usepackage{undertilde}
\usepackage[T1]{fontenc}
\usepackage[utf8]{inputenc}
\usepackage{comment}
\usepackage[export]{adjustbox}
\usepackage{scalerel}
\usepackage{mathtools}
\usepackage{leftindex}
\usepackage{tikz}
\usetikzlibrary{svg.path}

\definecolor{orcidlogocol}{HTML}{A6CE39}
\tikzset{
  orcidlogo/.pic={
    \fill[orcidlogocol] svg{M256,128c0,70.7-57.3,128-128,128C57.3,256,0,198.7,0,128C0,57.3,57.3,0,128,0C198.7,0,256,57.3,256,128z};
    \fill[white] svg{M86.3,186.2H70.9V79.1h15.4v48.4V186.2z}
                 svg{M108.9,79.1h41.6c39.6,0,57,28.3,57,53.6c0,27.5-21.5,53.6-56.8,53.6h-41.8V79.1z M124.3,172.4h24.5c34.9,0,42.9-26.5,42.9-39.7c0-21.5-13.7-39.7-43.7-39.7h-23.7V172.4z}
                 svg{M88.7,56.8c0,5.5-4.5,10.1-10.1,10.1c-5.6,0-10.1-4.6-10.1-10.1c0-5.6,4.5-10.1,10.1-10.1C84.2,46.7,88.7,51.3,88.7,56.8z};
  }
}

\newcommand\orcidicon[1]{\href{https://orcid.org/#1}{\mbox{\scalerel*{
\begin{tikzpicture}[yscale=-1,transform shape]
\pic{orcidlogo};
\end{tikzpicture}
}{|}}}}

\usepackage{hyperref}

\usepackage{subfiles}
\usepackage{xr}
\usepackage{siunitx}

\def\apj{ApJ}
\def\aap{A\& A}
\def\apjl{ApJL}
\def\apjs{ApJS}
\def\mnras{MNRAS}
\def\jcap{JCAP}
\def\pasa{PASA}

\def\aj{AJ}
\def\araa{ARA\& A}

\def\prd{Phys. Rev. D}

\def\rmxaa{RMxAA}

\def\pasp{PASPq}

\newcommand{\spec}{{\sc specwizard}}
\newcommand{\cloudy}{{\sc cloudy}}
\newcommand{\id}{{\sc id}}
\newcommand{\sph}{{\sc sph}}

\newcommand{\eagle}{{\sc eagle}}
\newcommand{\swift}{{\sc swift}}

\newcommand{\uvb}{{\sc uvb}}
\newcommand{\igm}{{\sc igm}}
\newcommand{\cgm}{{\sc cgm}}
\newcommand{\ism}{{\sc ism}}

\newcommand{\dla}{{\sc dla}}
\newcommand{\cddf}{{\sc cddf}}
\newcommand{\uves}{{\sc uves}}
\newcommand{\vlt}{{\sc vlt}}
\newcommand{\hires}{{\sc hires}}
\newcommand{\keck}{{\sc keck}}
\newcommand{\lya}{Lyman-$\alpha$}
\newcommand{\numpy}{{\sc numpy}}
\newcommand{\scipy}{{\sc scipy}}
\newcommand{\python}{{\sc python}}
\newcommand{\nist}{{\sc nist}}
\newcommand{\lls}{{\sc lls}}
\defcitealias{Theuns21}{TT21}
\newcommand{\kunit}{\hbox{km$^{-1}$~s}}
\newcommand{\vunit}{\hbox{km~s$^{-1}$}}

\DeclareRobustCommand{\ion}[2]{%
\relax\ifmmode
\ifx\testbx\f@series
{\mathbf{#1\,\mathsc{#2}}}\else
{\mathrm{#1\,\mathsc{#2}}}\fi
\else\textup{#1\,{\mdseries\textsc{#2}}}%
\fi}
\restylefloat{table}
\begin{document}

\title[\spec]{
\spec{}: A Public Python Framework for Creating Synthetic Absorption Spectra from Hydrodynamical Simulations}

\author[Arámburo-García, Theuns, Schaye \& Boyarsky]
{
 Andrés Arámburo-García$^{1, \orcidicon{0000-0003-4396-3491}}$, Tom Theuns $^{2, \orcidicon{0000-0002-3790-9520}}$, Joop Schaye$^{3, \orcidicon{0000-0002-0668-5560}}$ and Alexey Boyarsky$^{1, \orcidicon{0000-0003-0629-7119}}$\\
 $^{1}$ Institute Lorentz, Leiden University, Niels Bohrweg 2, Leiden, NL-2333 CA, The Netherlands\\
 $^{2}$ Institute for Computational Cosmology, Durham University, South Road, Durham DH1 3LE, UK\\
 $^{3}$ Leiden Observatory, Leiden University, PO Box 9513, NL-2300 RA Leiden, The Netherlands
}

\maketitle

\begin{abstract}
We present \spec{}, a publicly available modular Python package for generating synthetic absorption spectra from hydrodynamical simulations. The code combines atomic data from the {\em National Institute of Standards and Technology} database with ionisation fractions computed using the \cloudy{} spectral synthesis code for multiple ionising backgrounds, enabling consistent modelling of bound–bound and bound–free absorption for resonant transitions of all ions. \spec{} employs a column‑density–conserving projection scheme, includes natural line broadening for the \ion{H}{i} Lyman series, and offers optional treatments of instrumental resolution and noise for direct comparison with high‑resolution quasar spectra.
As example applications of the code, we shoot sight lines through the \eagle{} cosmological simulation to generate a $z=3.2$ synthetic quasar spectrum containing over $4\times 10^4$ \ion{H}{i} Lyman‑series transitions, including damping wings and Lyman‑limit absorption. Additionally, we compute metal-absorption lines probing the circumgalactic medium of simulated galaxies, and investigate the correlated absorption of \ion{H}{i} and \ion{Si}{iii}. We demonstrate that \spec{} applied to outputs from \eagle{} accurately reproduces key observational diagnostics of the Lyman‑$\alpha$ forest, including the redshift evolution of the mean transmission, the transmission {\sc pdf} and the transmission power spectrum. We take advantage of the fact that the \eagle{} simulation also reproduces the measured evolution of the temperature-density relation to infer the evolution of the amplitude of the ionizing background, $\Gamma$, finding excellent agreement with recent theoretical models and measurements.
\end{abstract}


\label{firstpage}

\section{Introduction}
\label{sec:Introduction}
The light from a redshift $z=3$ quasar travelled more than 11
billion light years before it hits our telescopes. An intervening structure may absorb a fraction of this light,
imprinting its presence in the quasar's spectrum. For example, when the sight line passes through a low-density region, the spectrum
will sample the low-density intergalactic medium (\igm). Sight lines passing close to a galaxy enable the study of the higher density gas near or even inside the galaxy (its circumgalactic, \cgm, and interstellar medium, \ism). This makes quasar spectra excellent probes of the distant Universe. The sight line also passes through the Milky Way's \cgm, allowing its study.

Examples of \igm\ studies include the determination of the ionisation and thermal state of the \igm\ and its enrichment with metals (ionisation state: e.g. \citealt{Bolton05, Bosman18, Ding24}; thermal state: e.g. \citealt{Schaye00, Rorai18, Hu23, Yang23}; enrichment: e.g. \citealt{Schaye03, Aguirre08}).

Examples of using quasars to study gas around galaxies and the ubiquity of galactic inflows or outflows include, e.g. \citealt{Adelberger05, Turner14, Turner17, Galbiati24}. For constraints on the nature of the Milky Way's \cgm, see e.g. \citealt{Zheng19}, and \citealt{Werk13} for the \cgm\ of other relatively nearby galaxies. Quasar spectra have also been used to constrain the nature of the dark matter \citep[e.g.][]{Viel13, Irsic17, Garzilli19, Irsic24} and of the dark energy, neutrino mass, and cosmological parameters 
\cite[e.g.][]{Busca13, Palanque15, Palanque20, Fernandez24}. The abundance of primordial Deuterium has also been determined \cite[e.g.][]{Cooke18}.

The interpretation of these quasar spectra typically requires comparing the observations with a model relating the detected features to the physical properties of the absorbing gas. Cosmological hydrodynamical simulations are now able to produce populations of {\em galaxies} with various degrees of realism (see for example \citealt{Schaye15} for \eagle\, \& \citealt{Schaye26} for {\sc colibre},  \citealt{Kaviraj17} for {\sc horizonAGN}, \citealt{Springel18} for {\sc IllustrisTNG}, \citealt{Dave19} for {\sc simba}, \citealt{Feldmann23} for {\sc firebox}). The extent to which these simulations also reproduce key observables of the \ism\ and \cgm\ of galaxies in {\em absorption} might put strong constraints on the realism of the simulations, and in particular, on the realism of the implemented feedback.

Several public codes are available to generate mock absorption spectra from simulations for comparison to data. The {\sc trident} code by \cite{Hummels17} is most similar in design and applicability to the \spec{} code described here, the {\sc LyaCoLoRe} code by \cite{Farr20} aims at generating survey-like spectra for comparison to baryon acoustic oscillation measurements.

In this paper, we present and verify the \textsc{python} implementation of 
\textsc{specwizard}\footnote{Source code and documentation: \url{https://github.com/specwizard/specwizard}; 
Python Package Index: \url{https://pypi.org/project/specwizard/}}, which is loosely based on an earlier {\sc fortran} code described by \citealt{Theuns98} and used by, for example, \cite{Schaye00} and \cite{Schaye03}. Our coding aims were to make the code easy to understand while sufficiently modular to enable users to adapt it to their needs. Implementation details are given in section~2. We use \spec\ to generate spectra from the \eagle\ simulation \citep{Schaye15, Crain15, McAlpine16} and compare them to selected observations in section~3. We use the good agreement between simulated and observed \lya-forest flux statistics to infer the value of the ionising background in section~4. We summarize in section~5.




\ifSubfilesClassLoaded{\bibliography{bibliography}{mybib}}

\end{document}

\section{The three steps for computing mock transmission spectra}
\label{sec:Method}
The \spec\ code computes the optical depth, $\tau$, due to scattering of photons from a background source by transitions from intervening ions in their electronic ground state (\lq resonant transitions\rq). This section describes the three steps \spec\ takes to calculate $\tau$ from the physical quantities calculated by the simulation. For Smoothed Particle Hydrodynamics (\textsc{sph}) simulations such as \textsc{gadget} 
\citep{Springel05} or {\sc swift}  \citep{Schaller24}, interpolation is performed  using the \textsc{sph} scheme itself; \textsc{swift} snapshots are read using the \textsc{swiftsimio} library \citep{Borr20swiftsimio}. This method would also be applicable to {\sc arepo} moving mesh \citep{Springel10} simulations. Mesh-based schemes, such as {\sc enzo} \citep{Bryan14} or {\sc ramses} \citep{Teyssier02}, should use the information from their grid directly.

Since a major motivation for developing \spec{} is to compare to data, it is worth briefly considering the difference between mock and observed spectra.
Observed spectra do not provide the optical depth directly. Instead, a reduced observational spectrum is the run of flux in wavelength bins, $F(\lambda_i)$, and an estimate of the noise per pixel in each bin, $n(\lambda_i)$. High-resolution Echelle spectra\footnote{With \lq high-resolution\rq\ we mean a spectral resolution of ${\cal R}\equiv \lambda/\Delta\lambda\gtrapprox 20, 000$, where $\Delta\lambda$ is the width of the line spread function.}, such as those from the \uves\ spectrograph on the \vlt\footnote{\url{https://www.eso.org/sci/facilities/paranal/instruments/uves.htm}}
or \hires\ on \keck\footnote{\url{https://www2.keck.hawaii.edu/inst/hires/}},
can be wavelength calibrated very accurately, however, it is much harder to calibrate the flux
\citep[e.g.][]{Suzuki03, Brahm17}. The function $F(\lambda_i)$ therefore depends on the properties of the instrument (in particular on the wavelength sensitivity of the detector), on the shape of the 
background source (typically a quasar), i.e. of a hypothetical spectrum with no intervening absorption, in addition to any signatures of intrinsic and intervening absorption. The step in the data reduction from observed flux to inferred transmission, $T=F/F_{\rm cont}\equiv \exp(-\tau)$, where $F_{\rm cont}$ is the flux that would have been observed in the absence of absorption (\lq continuum fitting\rq) is not modelled in \spec, however, it does include models to account for the detector's thermal noise and the spectrum's photon noise.

\spec\ computes absorption by electronic transitions of ions as a function of a \lq Doppler\rq\ velocity $v$, defined by,
\begin{align}
    \frac{dv}{c}=\frac{d\lambda}{\lambda}\,;\quad \frac{v}{c}=\ln(\frac{\lambda}{\lambda'})\,,
\end{align}
where $\lambda$ is observed wavelength and $\lambda'$ the (arbitrary) observed wavelength where $v=0$. Using $v$ rather than $\lambda$ makes it easy to include peculiar velocities, Hubble flow, and physical line broadening. Setting $\lambda'=\lambda_0(1+z)$ for a given ion with laboratory wavelength $\lambda_0$ at redshift $z$ allows us to easily compare the lines of different ions that are produced by the same intervening gas cloud. The spectral resolution of spectrographs, ${\cal R}=\lambda/\Delta\lambda$, is often approximately constant, in which case the spectral resolution in velocity space, $\Delta v$, is also approximately constant. Finally, we note that $v$ is simply a change of variables from $\lambda$, so there are no special relativistic corrections and no requirement for $v$ to be smaller than $c$.

Given a sight line through the simulation volume, divided in $N$ pixels,
the three steps to compute $\tau(v_i)$ for each pixel $i$ with Doppler velocity $v_i$, are
\begin{enumerate}
    \item[(1)] {\em ionisation balance}: determine the fraction $X\equiv n_{\rm ion}/n_{\rm element}$ for each particle and for each ion; $n_{\rm ion}$ is the ionic number density of ion $X$, and $n_{\rm element}$ is the number density of the corresponding element,
    \item[(2)] {\em particle projection}: obtain the ion column density for each pixel by integrating the ion density for all particles and ions,
    \item[(3)] {\em optical depth calculation}: compute the optical depth
    for each ion in each pixel due to the ionic column density in all pixels.
\end{enumerate}

For ions with more than one transition, \spec{} performs these steps for the transition with the largest oscillator strength, and the results are
saved for the sight line: we refer to this output as a \lq sight line\rq\ or \lq short spectrum\rq. In addition, \spec\ will compute and save optical depth-weighted averages of physical quantities such as temperature and density. Such short spectra can be used to examine correlations between lines of different ions in the same gas cloud, or study the relations between optical depth and physical properties of the gas responsible for the absorption. In the second step, optical depths for all other transitions of ions are obtained by scaling the saved optical depth by the ratio of line strengths\footnote{We use a different method for the Hydrogen atom to account for other Lyman series lines, as described below.}. This is possible because the ion fraction and line width is the same for all transitions of a given ion. In an optional third step, \spec\ combines short spectra into a mock observed spectrum, with the user selecting the redshift of the source, and the minimum and maximum \lq observed\rq\ wavelength. The code will then combine short spectra into one mock observed spectrum (total optical depth due to all transitions of all ions combined as a function of wavelength). This final step accounts for contamination (ions at different intervening redshifts producing absorption at the same observed wavelength) and also accounts for bound-free absorption, and natural broadening for \ion{H}{i} lines. The next section provides more details of the three basic steps.

\subsection{Step 1: ionisation balance}
\begin{figure}
 \includegraphics[width=\columnwidth]{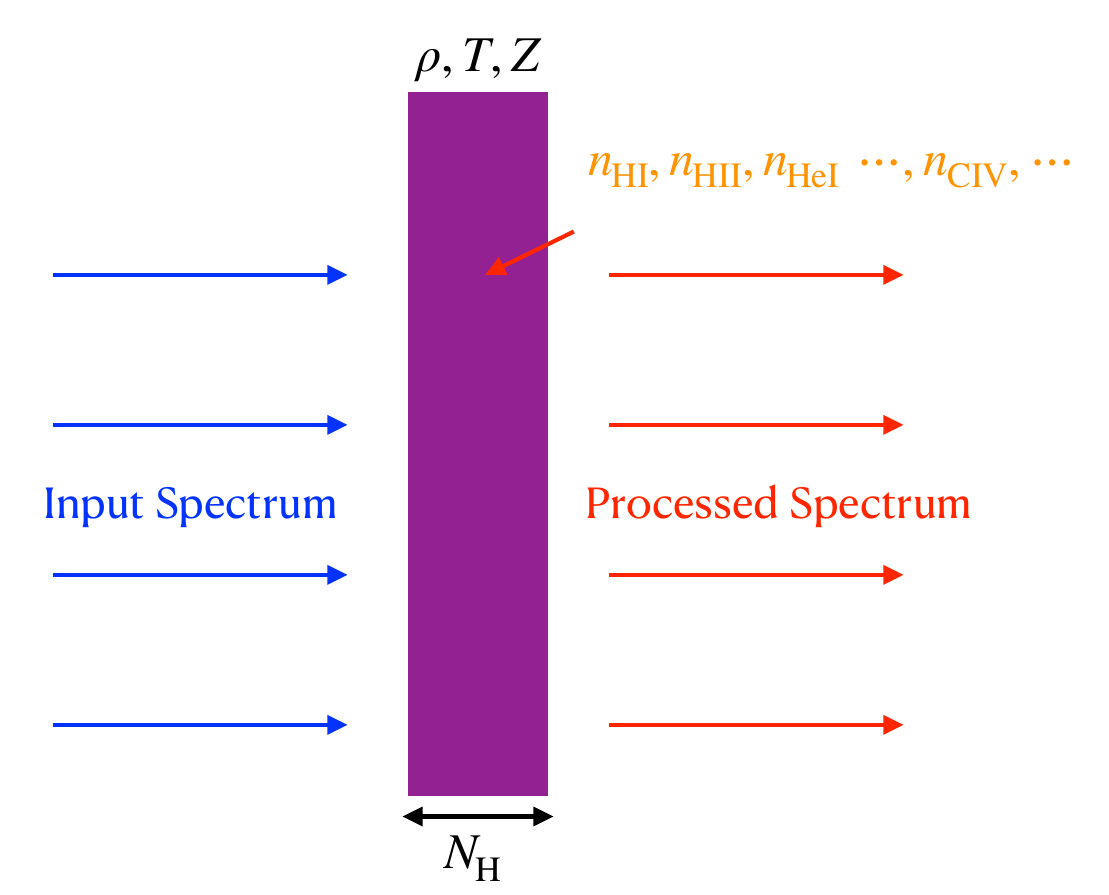}
\caption{Illustration of the \cloudy\ set-up.}
\label{fig:cloudy}
\end{figure}

\begin{figure}
 \includegraphics[width=\columnwidth]{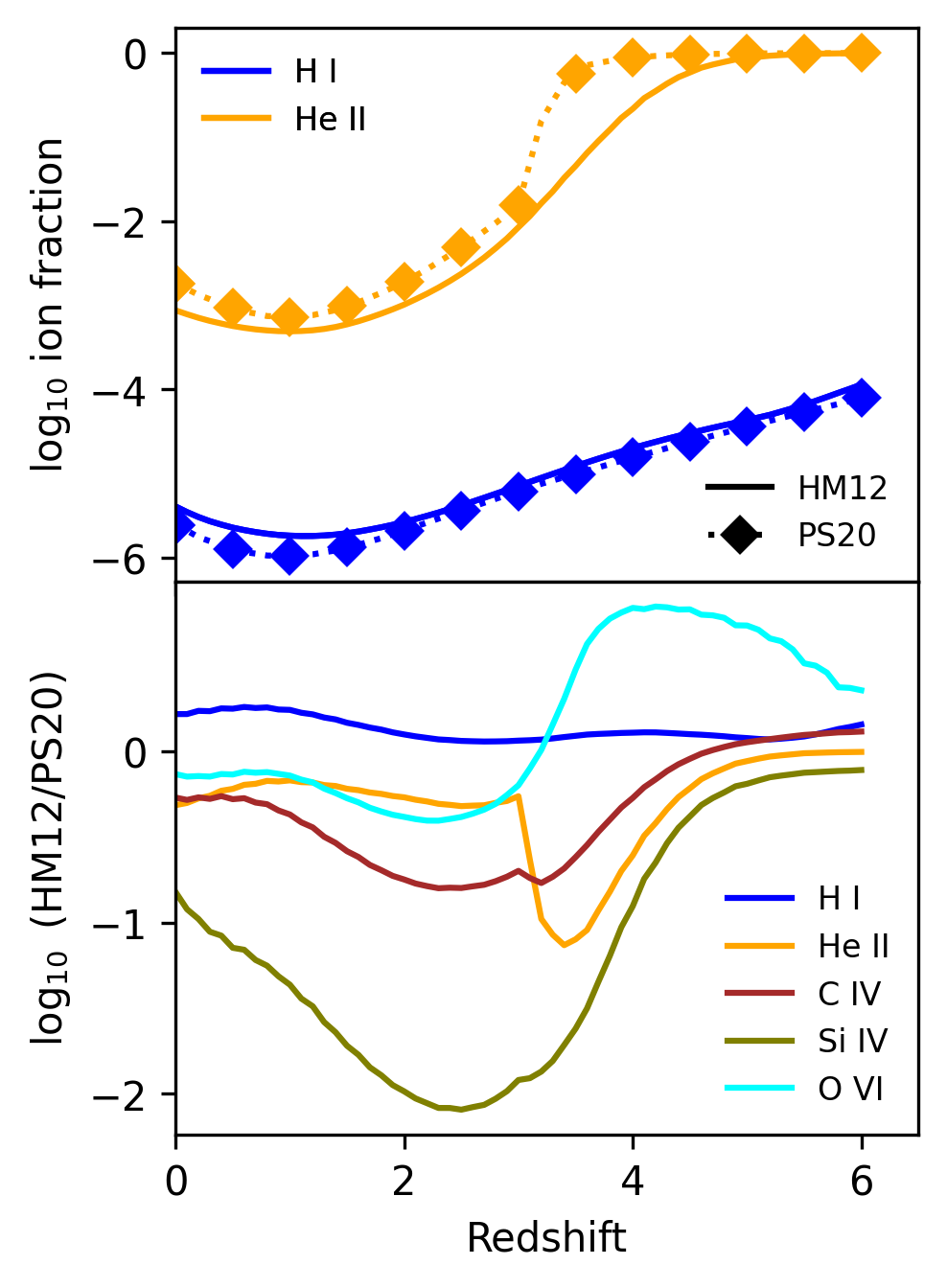}
\caption{{\em Upper panel}:  ionisation fraction as a function of redshift, $z$, for ions \ion{H}{i} ({\em blue}) and \ion{He}{ii} ({\em yellow}). {\em Solid lines} use the \lq HM12\rq\ ionisation tables computed using the ionising background computed by \protect\cite{Haar12UVbackground},  {\em dotted lines connecting diamonds} use the \lq PS20\rq\ tables computed by \protect\cite{Ploeckinger20} which use a modified version of the ionising background of \protect\cite{Faucher20}. In all cases, the gas is at the cosmic mean density and is at a temperature of $T=10^4{\rm K}$.  The sharpish feature around $z=3.5$ in the \protect\cite{Faucher20} calculation is due to the onset of \ion{He}{ii} reionisation. {\em Lower panel:} ratio of ionisation levels for these two tables for the ions in the legend. The different choices of ionising background have a profound impact on the ionisation level, with for example the \ion{Si}{iv} ionisation fraction 
at $z=2$ more than two orders of magnitude lower when using the \protect\cite{Haar12UVbackground} ionising background compared to the one computed by \protect\cite{Faucher20}.}
\label{fig:ionbal}
\end{figure}

We construct tables in the {\sc hdf5} format\footnote{Details of the hdf5 format can be found at \href{https://www.hdfgroup.org/solutions/hdf5/}{https://www.hdfgroup.org/solutions/hdf5/}.} to interpolate the
ratio of the ion density over the element density, $X\equiv n_{\rm ion}/n_{\rm element}$, for all ions, as a function of density and temperature (in logarithmic bins) and redshift (in linear bins). The calculation of $X$ is performed using the \cloudy{} code (version 23, \citealt{Cloudy23, Gunasekera23}), which solves the equations of statistical equilibrium and the equations that balance ionisation-neutralization processes for an astrophysical plasma irradiated by an external radiation field. In our case, the radiation field is an ambient background of ionising radiation from galaxies and quasars, and several choices for this redshift-dependent radiation field are implemented in \cloudy. These include the evolution computed by \cite{Fauc09}, \cite{Haar12UVbackground} \& \cite{Faucher20}. The \python\ wrappers that generate the required \cloudy\ input files, and those that parse the \cloudy\ output are part of the \spec\ distribution. These wrappers can be modified if the user prefers a different radiation field, or wants to add or remove elements/ions. \spec\ also allows the user to use published ionisation tables, for example, those of \citealt{Ploeckinger20}, which account for self-shielding. 

The \cloudy\ set up is that of plane-parallel slab of gas with specified values of the hydrogen density ($n_{\rm H}$), gas temperature ($T$), and chemical composition (\lq metallicity\rq), as illustrated in Fig.~\ref{fig:cloudy}. The thickness of the slab is $1~\unit{cm}$, ensuring that the calculation is in the optically thin limit. The default set up uses \cloudy's solar chemical abundance pattern, but with metallicity reduced by a factor of $10^{4}$ except for the helium abundance which is set to its primordial value. Such a low metallicity ensures that any electrons released from metals have little impact on the ionisation state of ions, and hence ionic abundances are to a good approximation simply proportional to the abundance of the corresponding element. 
\spec{} then computes the ion abundance by interpolating the computed tables. For example, the ion density for \ion{H}{i} for a particle is
\begin{align}
    n_{{\rm HI}}(\rho, T, z) &= x_{{\rm HI}}(\rho, T, z)\,X_{\rm H}\,\rho/m_{\rm H}\,,
    \label{eq:xHI}
\end{align}
where $X_{\rm H}$ is the hydrogen fraction by mass of the particle, $m_{\rm H}$ the mass of the hydrogen atom, and $x_{{\rm HI}}(\rho, T, z)$ the \ion{H}{i}
fraction for gas with density $n_{\rm H}$ and temperature $T$ at redshift $z$, as interpolated from the tables.

The importance of the choice of ionizing background is illustrated by Fig.~\ref{fig:ionbal}, where we plot the ionisation state of gas at the mean cosmic density with a temperature of $10^4{\rm K}$ as a function of redshift for different ions. We compare our \cloudy{} results for the \cite{Haar12UVbackground} ionising background against the tables of \cite{Ploeckinger20}, that use a modified version of the ionising background of \cite{Faucher20}. The \ion{H}{i} fractions computed using
the two choices of background are very similar, but the onset of \ion{He}{ii} reionisation is more sudden in the latter table, as can be seen by the sudden drop in the \ion{He}{ii} fraction around $z\approx 3.5$ in the 
\cite{Ploeckinger20} abundance which is not present in the \cite{Haar12UVbackground} case (see the upper panel).  The two ionising backgrounds result in significant differences, by up to two {\em orders of magnitude}, in the ionisation states of several commonly observed ions, such as \ion{Si}{iv} and \ion{O}{vi} (lower panel). 

Some simulations track ion abundances explicitly, using radiative transfer computed either on the fly \citep[e.g.][]{Pawlik08, Rosdahl18, Kannan19,Hopkins20, Chan21, Chan24} or in post-processing \citep[e.g.][]{GIZMO, Alta13revrt}, thereby accounting for self-shielding and/or non-equilibrium ionisation that the optically thin \cloudy{}  tables neglect. In this case, 
\spec\ can read the ion fraction of each particle directly from the snapshot in place of the tabulated value in Eq. (\ref{eq:xHI}), leaving the subsequent steps unchanged, and computes two sets of spectra, one using the \cloudy{}  tables and one using the simulation's own fractions, so that the impact of these effects can be assessed directly.

\subsection{Step 2: Particle projection}
\label{sect:projection}
\begin{figure}
 \includegraphics[width=\columnwidth]{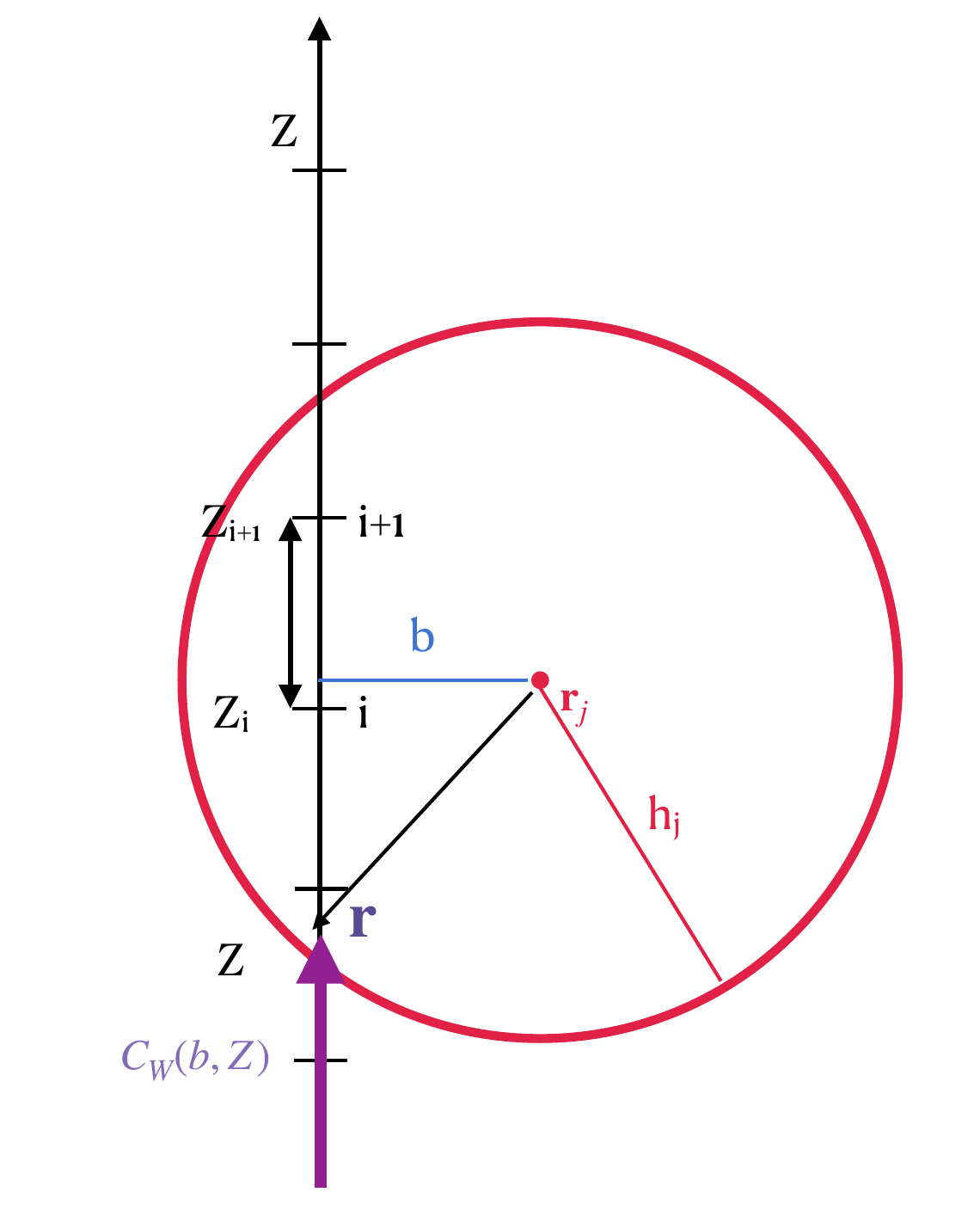}
\caption{Cartoon of a line of sight ({\em black vertical line}) intersecting an {\sc sph} particle ({\em red circle}) that is located at position ${\bf r}_j$ and has smoothing length $h_j$. $Z_i$ is a pixel along the sightline, which runs from coordinate $Z_i\to Z_{i+1}$. To implement the calculation of column densities, we compute the cumulative column density from $Z=-\infty$ up to the end of each pixel, Eq.~(\ref{eq:CW}), as illustrated by the {\em purple} line labelled $C_W(b,Z)$.
The column density for pixel $i$ is $C_W(b,Z_{i+1})-C_W(b, Z_i)$.}
\label{fig:projection}
\end{figure}
\begin{figure}
 \includegraphics[width=\columnwidth]{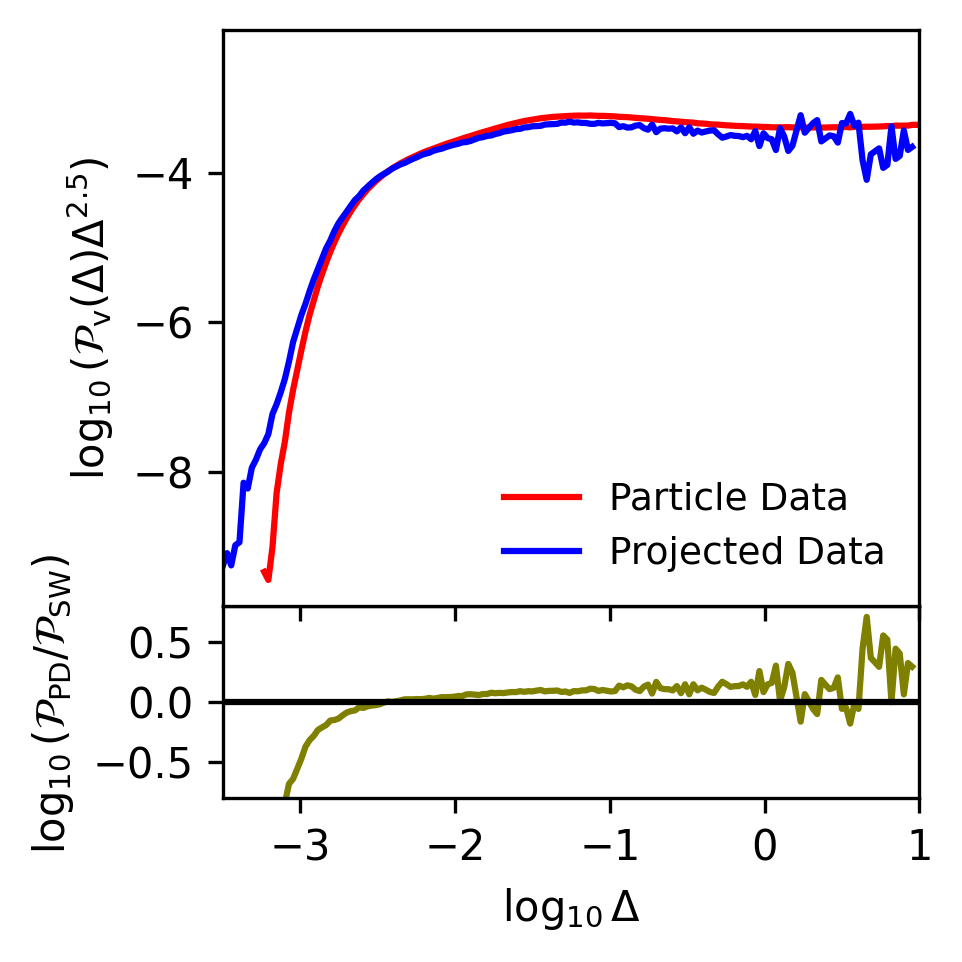}
\caption{Upper panel: The probability distribution function of overdensity by volume, $\mathcal{P}_{\mathrm{v}}(\Delta)$, 
computed directly from all the particles ($\mathcal{P}_{\mathrm{v}, PD}$, {\em red line}, Eq.~\ref{eq:pVp}) and from the spectra ($\mathcal{P}_{\mathrm{v}, S}$, {\em blue line}, Eq.\ref{eq:pVsw}) for the \eagle\ simulation {\sc recal}L0050N0752. Both PDFs are multiplied by $\Delta^{2.5}$ to decrease the dynamic range. {\em Lower panel}: ratio of the PDFs. At the lowest densities, $\mathcal{P}_{\mathrm{v}, SW}>\mathcal{P}_{\mathrm{v}, \rm PD}$, whereas at the densities around the mean, $\log_{10}\Delta\in[-1,1]$), $\mathcal{P}_{\mathrm{v}, PD}$ is about 10~per cent higher than $\mathcal{P}_{\mathrm{v}, SW}$. The calculation using spectra suffers from sample variance at the highest densities
(we used 500 spectra). The relatively good agreement between the two methods demonstrates that the projection scheme samples the \sph\ density field faithfully.}
\label{fig:projection2}
\end{figure}

Particle projection refers to the step of computing column densities for ions for each pixel of a sight line from the ion densities of particles, as computed in the previous step; ion density-weighted peculiar velocities and temperatures are computed as well. \spec{} uses the {\sc sph} interpolation scheme to project particles, but in such a way as to ensure that the column density is conserved - independent of pixel size. The method takes advantage of the efficient implementation of interpolation in the {\sc numpy} \python\ library.

The interpolation scheme is illustrated in Fig.~\ref{fig:projection}. The vertical line is the chosen sight line, which intersects a simulation particle (shown as a circle with radius equal to the smoothing length); short horizontal lines indicate the pixels along the sight line. We want to compute the ionic column density due to this particle in each of the pixels that intersect the particle. We accomplish this as follows.

The \sph\ interpolation scheme in the gather interpretation allows us to calculate the contribution to the density at any location ${\bf r}$ from a particle $j$ with mass $m_j$ at position ${\bf r}_j$ as
\begin{align}
    \rho({\bf r},{\bf r}_j)=& m_j\,W(q,h_j)\,;\quad q\equiv \frac{|{\bf r}-{\bf r}_j|}{h_j}\,,
\end{align}
and the neutral hydrogen density by number as
\begin{align}
    n_{\rm HI}({\bf r},{\bf r}_j)= x_{{\rm HI}}(\rho, T, z)X_{\rm H}\,\rho({\bf r}, {\bf r}_j)/m_{\rm H}\,,
\end{align}
where we use the ion abundance computed as in Eq.~(\ref{eq:xHI}). We use the {\sc m4} spline, 

\begin{equation}
W(q,h)=
    \begin{cases}
        ({8}/{(\pi h^3}))\,\left(1-6q^2+6q^3\right)\ & \hbox{\rm if } q < 1/2\\
        ({8}/{(\pi h^3}))\,2(1-q)^3 & \hbox{\rm if } 1/2<q<1\,,
    \end{cases}
\label{eq:kernel}
\end{equation}
where $h$ is the smoothing length of the particle, although it would be easy to use any other kernel\footnote{In this notation, $W(q,h)=0$ for $q>1$, i.e. $h$ is the radius of the kernel's compact support. The meaning of $h$ is different in the \swift{} code.}. The contribution to the total column density of this particle is an integral of $W$ at a given impact parameter, $b$. We ensure that this column density is calculated accurately by computing and tabulating the integral
\begin{align}
    {\cal C}_W(b, Z) &\equiv \int_{-\infty}^Z\,W\left((Z'^2+b^2)^{1/2}, h=1\right)\,dZ'\,,
    \label{eq:CW}
\end{align}
as a function of $Z$ and impact parameter $b$, on a fine mesh of $Z$ and $b$ values (i.e. with step size $|dZ|$ and $|db| = 0.01$). We scale this table to use the actual value of the smoothing length $h_j$ of the particle and use linear interpolation to find the cumulative column density for each pixel $i$. Differencing the cumulative column then yields the required weight $\int_{Z_i}^{Z_{i+1}}\,W(q, h_j)\,dZ$. We refer to this {\bf I}nterpolation followed by {\bf D}ifferencing operation as \id, and denote it by ${\cal D}$. The neutral hydrogen column density associated with pixel $i$ is then
\begin{align}
    N_{\rm HI}({\bf r}_i, {\bf r}_j) &= \frac{x_{{\rm HI},j}\,X_{{\rm H}, j}}{m_{\rm H}}\,m_j\,\int_{Z_i}^{Z_{i+1}}\,W\left({\frac{|{\bf r}'_i-{\bf r}_j|}{h_j}}, h_j\right)\,dZ'\nonumber\\
    &\equiv \frac{x_{{\rm HI},j}\,X_{{\rm H}, j}}{m_{\rm H}}\,m_j\,{\cal D}_{i,j}\left[W\right]\,.
    \label{eq:NHI}
\end{align}
This method of interpolation differs from that by \cite{Theu98P3MSPH}, who computed column densities more approximately as
$N_{\rm HI}({\bf r}_i, {\bf r}_j)\approx n_{\rm HI}({\bf r}_i, {\bf r}_j)\,(Z_{i+1}-Z_i)$.  

Summing over all particles yields the total column density, and we similarly compute the ion-density-weighted temperature, $T_{\rm HI}$, and peculiar velocity, $v_{Z, {\rm HI}}$, as
\begin{align}
    N_{\rm HI}({\bf r}_i) &= \sum_j \int_{Z_i}^{Z_{i+1}}\,
    n_{\rm HI}({\bf r}'_i, {\bf r}_j)\,
    dZ'\nonumber\\
    T_{\rm HI}({\bf r_i}) &= \sum_j \int_{Z_i}^{Z_{i+1}}\,
    n_{\rm HI}({\bf r}'_i,{\bf r}_j)\,
    T_{\rm j}\, dZ'\,/N_{\rm HI}({\bf r}_i)\nonumber\\
    v_{Z,{\rm HI}}({\bf r_i}) &= \sum_j \int_{Z_i}^{Z_{i+1}}\,
    n_{\rm HI}({\bf r}'_i,{\bf r}_j)\,
    v_{Z, {\rm j}}\, dZ'\,/ N_{\rm HI}({\bf r}_i)\,.
    \label{eq:projection}
\end{align}
Here, $T_j$ and $v_{Z,j}$ are the temperature and peculiar velocity along the sight line of particle $j$. We evaluate Eq.~(\ref{eq:projection}) for each ion to obtain the ion column density, ion-weighted temperature, and ion-weighted peculiar velocity. Although column density is conserved by construction, we still need to ensure that $dZ$ is sufficiently small compared to $h_i$ for all particles that contribute to the sight line to avoid loss of numerical accuracy.

As an illustrative example of the interpolation scheme, we compute the probability distribution of overdensities by volume, $\mathcal{P}_V(\Delta)$, in two ways. Using the \sph\ densities for each particle, we evaluate $\mathcal{P}_V(\Delta)$ as 
\begin{align}
    \mathcal{P}_{V, {\rm PD}}\, d\Delta &=
    \sum_{i\in d\Delta}\rho^{-1}_i/
    \sum_{i=1}^{N} \rho^{-1}_i\,,
    \label{eq:pVp}
\end{align}
where $\rho_i$ is the \sph\ density of particle $i$. The numerator is the sum of the inverse of the densities of all particles in a given over density bin, $d\Delta$, the denominator is the normalization. Shooting many sight lines through the same simulation, we alternatively compute $\mathcal{P}_V(\Delta)$
as 
\begin{align}
    \mathcal{P}_{v, {\rm SW}}\, d\Delta &=
    \sum_{s\in dN} / N_{\rm pix}\,,
    \label{eq:pVsw}
\end{align}
The numerator is the number of pixels in the sight line where the projected density falls inside a given over density bin, $d\Delta$, the denominator is the normalization.

Figure~\ref{fig:projection2} compares the two methods. For over densities around the mean, the two methods agree to within 10 per~cent. At high over densities, $\log_{10}\Delta\gtrapprox 1$, the method that uses sight lines suffers from sample variance due to the limited number of sight lines computed, whereas at low over densities, $\log_{10}\Delta\lessapprox -2.5$, the particle method falls below the sight line estimate. We note that the two methods compute different quantities and there is no reason that the two should give exactly the same answer. We also note that the projection method conserves mass by construction.

\subsection{Step 3: Optical depth calculation}
In this final third step, we use the column densities computed for each pixel in the previous step to determine the optical depth, $\tau$, in all pixels. We illustrate the method for transitions in hydrogen; other elements are discussed in the next section. The implementation ensures conservation of equivalent width, independently of pixel size. We begin by recalling how $\tau$ and column density are related for a slab. Interpreting each pixel as a slab then shows how to compute $\tau$ for each pixel in the sight line.

\label{sec:OpticalDepth}
\subsubsection{Bound-bound transitions}
Consider a thin slab of gas with co-moving thickness $dZ$ that is perpendicular to the sight line at co-moving position $Z_i$, and assume that the co-moving \ion{H}{i} density is $n_{\rm HI}(Z)$, the \ion{H}{i}-weighted temperature is $T$, and the \ion{H}{i}-weighted peculiar velocity along the sight line direction is $v_{{\rm p}}$: these quantities are computed as described in the previous section. The Hubble velocity of the slab is $v_{{\rm H}}=Z\,H(z)/(1+z)$, and the total velocity is $v=v_{{\rm H}}+v_{{\rm p}}$. When a beam of light along the line of sight strikes the slab perpendicular to it, some of its photons may excite hydrogen atoms from the ground state $n=1$ to the first excited state\footnote{We use the usual notation where $n=1$, 2, $\cdots$ is the principal quantum number and $l=0,\cdots, n-1$ is the angular momentum quantum number.}, leading to the absorption of Lyman-$\alpha$ photons. The excited electrons will fall back to the ground state and hence re-emit Lyman-$\alpha$ photons but generally in a different direction. The process of absorption and re-emission is therefore better thought of as the scattering of Lyman-$\alpha$ photons, but we will follow convention and refer to it as absorption. 

The fraction of light that is absorbed by the slab at velocity $v'$ is $\exp(-d\tau(v'))$, with the optical depth given by
\begin{align}
    d\tau(v') = \sigma\,c\,\big\{(1+z)^3\,n_{\rm ion}(Z)\big\}\,\Phi(v'-v)\,\frac{dZ}{1+z}\,,
    \label{eq:dtau}
\end{align}
see e.g. \cite{Rybicki86} p.~287. The factor $(1+z)^3$ converts from co-moving to proper density, and $dZ/(1+z)$ is the proper thickness of the slab; $c$ is the speed of light. The cross section $\sigma$ is given by
\begin{align}
    \sigma &=  \left(\frac{3\pi\sigma_{\rm T}}{8}\right)^{1/2}\,\lambda_0\,f\,,
    \label{eq:sigma}
\end{align}
where $\sigma_{\rm T}$ is the Thomson cross section, and $\lambda_0$ and $f$ are the laboratory (vacuum) wavelength\footnote{Historically, optical wavelengths 
were quoted in air, but the recent trend is to quote all wavelengths in vacuum.} of the transition and the oscillator strength. As an example, for the \lya{} transition, $\lambda_0\equiv \lambda_2\approx 1215.67$\AA~ and $f\equiv f_2\approx 0.4164$, yielding $\sigma\approx 4.48\times 10^{-18}\unit{cm^{2}}$. The
expression for $d\tau$ differs from that in \cite{Rybicki86} because we write $d\tau$ in terms of velocity rather than frequency and this introduces the factor $\lambda_0=dv/d\nu$. $\Phi$ is the fraction of ions that move with speed $v$, normalised to $\int_{-\infty}^\infty \Phi(v)\,{\rm d}v=1$ - this function describes the line shape. If the velocity distribution were purely thermal, then $\Phi=\Phi_{\rm G}$, where
\begin{align}
    \Phi_{\rm G}(v, T) &= \frac{1}{(\pi\,b_D^2)^{1/2}}\,\exp(-\frac{v^2}{b_D^2})\,;\quad
    b_D^2(T)= \frac{2k_{\rm B}T}{m}\,,
\label{eq:gauss}
\end{align}
where $k_{\rm B}$ is Boltzmann's constant, $T$ the ion-weighted gas temperature, and $m$ the element's mass\footnote{The Doppler parameter, $b_D$, used here is not to be confused with the impact parameter used in the previous section.}.

In the special case of a single, thin slab at location $Z$ with velocity $v$,  
Eq.~(\ref{eq:dtau}) can be simplified by integrating the density over the slab, 
\begin{align}
    \int_{-\infty}^\infty (1+z)^3\,n_{\rm ion}\,\frac{dZ}{1+z} &\equiv N_{\rm ion}\,,
\end{align}
where $N_{\rm ion}$ is the (proper) column density. The optical depth profile of ion \ion{H}{i} is then
\begin{align}
\tau(v') &= \sigma\,c\,N_{\rm HI}\,\Phi(v'-v)\,,
\end{align}
which is a Gaussian for a thermal distribution of velocities (i.e. when $\Phi=\Phi_{\rm G}$). The total optical depth due to many slabs follows from summing over slabs,
\begin{align}
    \tau(v') &= \sigma\,c\,\sum_i N_{{\rm ion},i}\,\Phi(v'-v_i)\,.
\label{eq:tausum}
\end{align}
We showed in the previous section how we use the {\sc sph} particles to compute the column density and ion-weighted peculiar velocity per pixel. The previous equation is then a useful way to conceptualise how the spectrum should be computed: every pixel (with a given value of column density and velocity) contributes to the optical depth in every other pixel, by the amount given by
Eq.~(\ref{eq:dtau}). 

It is also possible to integrate Eq.~(\ref{eq:dtau}) along the line of sight to obtain
\begin{align}
    \tau(v')&=\sigma\,c \int_{-\infty}^{\infty}\,(1+z)^2\,n_{\rm HI}(Z_i)\,\Phi\left(v'-v_i\right)\,dZ\nonumber\\
                   &=\sigma\,c\,\frac{dN_{\rm HI}(v')}{dv'}
    \label{eq:tgauss}               
                   \\
    \frac{dN_{\rm HI}(v)}{dv} &\equiv \int_{-\infty}^{\infty}\,\frac{(1+z)^2}{|dv'/dZ|}\,n_{\rm HI}(v')\,\Phi\left(v'-v\right)\,dv'\,,
    \label{eq:dNdv}
\end{align}
a result sometimes referred to as the \lq fluctuating Gunn-Peterson\rq\ approximation
\citep{Croft98, Gnedin98}. The second line shows that the optical depth is the product of the number of ions per unit area and per unit velocity, $dN_{\rm HI}(v)/dv$, times $\sigma\,c$. 

Given the column density in pixel $i$ as calculated in the previous section, we could use Eq.~(\ref{eq:tausum}) to compute the corresponding optical depth. 
The presence of the velocity profile $\Phi$ - typically taken to be the Gaussian profile of Eq.~(\ref{eq:gauss}) - requires that the pixel size ($dv$) used to perform the integration is sufficiently small compared to the parameter $b_D$ for each pixel. In fact, the expression for $\tau$ shows that what is required, is that the integral of $\tau$ over a pixel is accurately computed - since this is the integral of $dN_{\rm ion}/dv$, i.e. the total number of ions that contributes to the optical depth in that pixel. The values of $b_D$ for hydrogen are typically larger than $\approx 10~{\rm km~s}^{-1}$ or more, but those of heavier ions (e.g. those of oxygen or iron) can be significantly narrower since $b_D^2$ is inversely proportional to the mass of the element. We note that even in case we want to compare to a spectrum in which these lines are not actually resolved, our mock spectrum must be accurately computed since otherwise our realization of the mock spectrum does not have the correct equivalent width.

We ensure that the equivalent width is conserved  by using the \id\ method of \S \ref{sect:projection} on the function $\Phi_{\rm G}$: we compute and tabulate the cumulative integral of the Gaussian of Eq.~(\ref{eq:gauss}) which appears in Eq.~(\ref{eq:dNdv}) 
\begin{align}
    {\cal C}_{\Phi_{\rm G}}(v) = \int_{-\infty}^v\,\Phi_{\rm G}(v',b_D=1)\,dv'\,,
\end{align}
- i.e. we tabulate the error function - and then scale the tabulated function to account for the actual value of the line-broadening parameter, $b_D$. As in the case of particle projections, the \python\ implementation of this calculation takes advantage of the efficient implementation of interpolation in \numpy.

Approximating lines as Gaussian is usually a good approximation but fails when the optical depth at the line centre becomes large. In that case, we need to account for the intrinsic line width of the transition (as opposed to its thermal width): the intrinsic width of lines gives rise to \lq damped \lya-absorbers\rq\ (\dla s). We also need to include the effect of bound-free transitions (i.e. ionisations), which give rise to Lyman-limit systems (\lls s). We discuss the corresponding line shapes next.

\subsubsection{Natural line widths: \lq damping wings\rq}
\begin{figure*}
     \includegraphics[width=\textwidth]{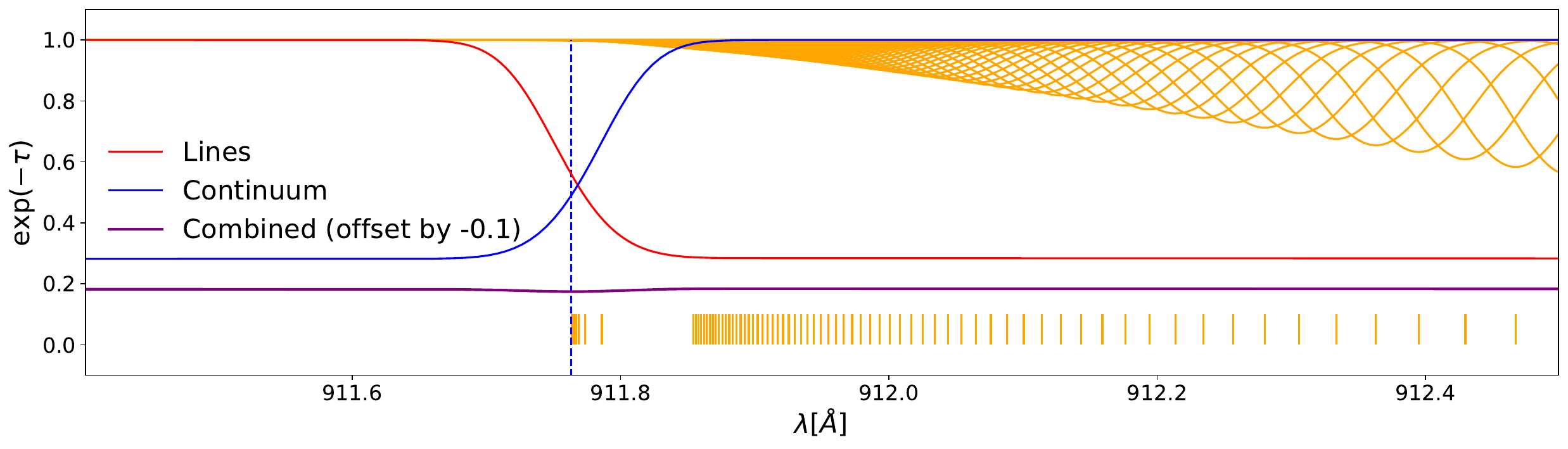}
      \caption{Example spectrum of an \ion{H}{i} absorber with column density $N_{\rm HI}=10^{17.3}{\rm cm}^{-2}$ and thermal and turbulent broadening parameters of $12~{\rm km~s}^{-1}$, as computed using the convolution method. The calculation accounts for 40,000 Lyman series lines, with individual line profiles plotted in {\em orange}, with {\em short orange vertical lines} indicating their central wavelength. All first 100 lines are shown, then we plot only every hundredth Lyman line up to Lyman 2,000. The {\em red line} is the net transmission of all 40,000 lines combined. The {\em blue line} is due to bound-free absorption, with the {\em vertical blue dashed line} indicating the wavelength corresponding to ionisation. Thermal and turbulent broadening cause bound-free absorption to extend to the red side of the ionisation threshold, and line absorption to its blue side. The net profile is the combination of line and bound-free absorption, and is plotted as a {\em purple line}; the total transmission is offset vertically by -0.1 for clarity. The combination of line and bound-free absorption make the total optical depth independent of wavelength close to the ionisation edge.}
\label{fig:edge}
\end{figure*}

\begin{figure*}
     \includegraphics[width=\textwidth]{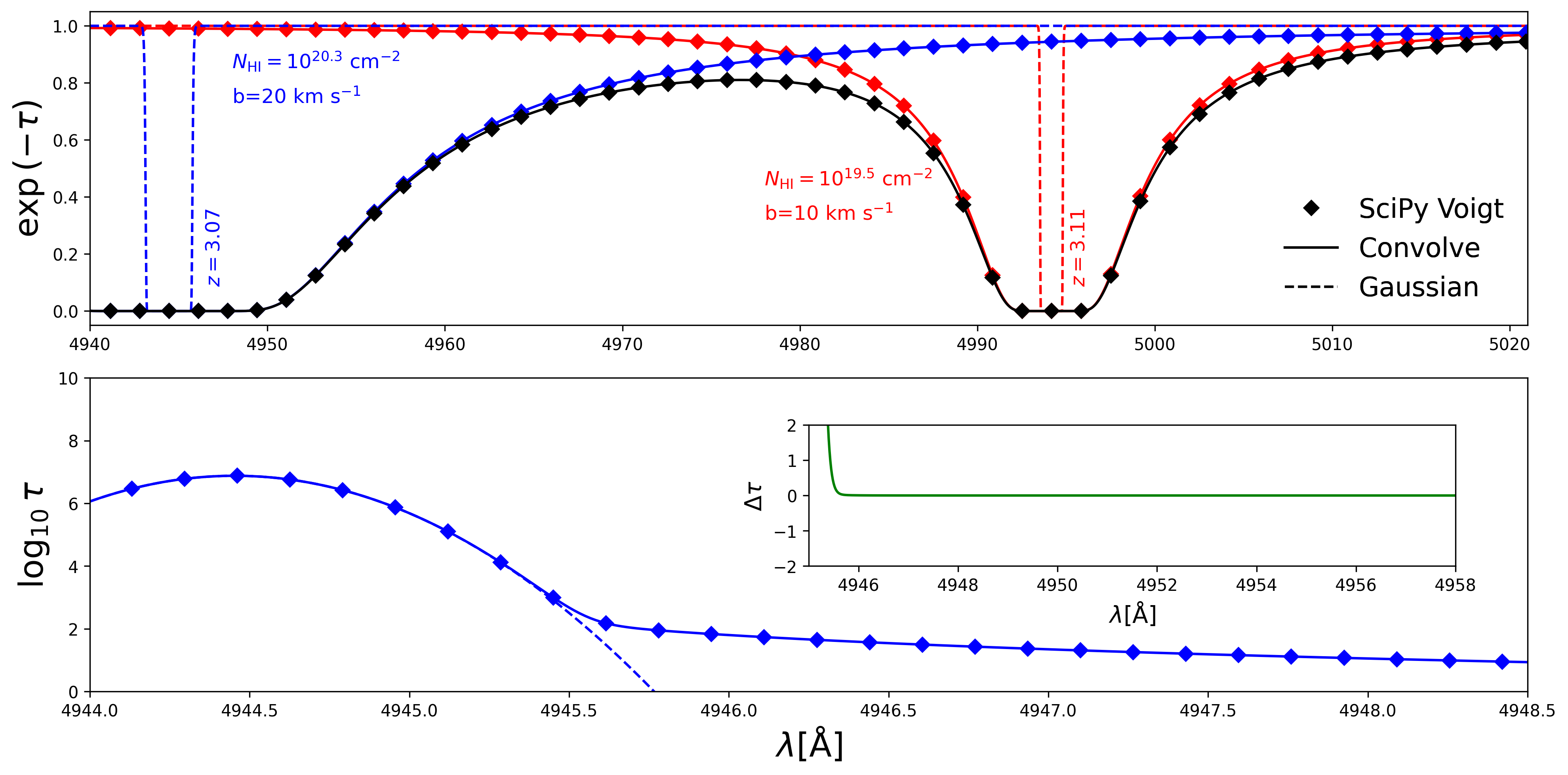}
     \caption{Illustration of the implementation of the  \ion{H}{i} damping wing. The spectrum contains two \dla s with parameters $N_{\rm HI}=10^{20.3}~{\rm cm}^{-2}$, $b_D=20~{\rm km~s}^{-1}$ and $z=3.07$ and $N_{\rm HI}=10^{19.5}~{\rm cm}^{-2}$, $b_D=10~{\rm km~s}^{-1}$ and $z=3.11$. The column density of all other pixels is set to zero. In the  {\em lower panel}, we plot optical depth as a function of wavelength for the stronger \dla, comparing our implementation using Eq.~(\ref{eq:damping}) ({\em solid blue line}) to the profile computed with the \scipy\ function {\sc voigt\_profile} ({\em blue diamonds}); the {\em dashed blue line} is the Gaussian profile for this absorber. The {\em small inset} shows the difference in optical depth, $\Delta \tau$: the relative error in $\Delta\tau$ is small everywhere, even at the line centre. The {\em upper panel} shows the case of the two \dla s. The {\em red} and {\em blue lines and symbols} correspond to the individual profiles, the {\em black line} shows Eq.~(\ref{eq:damping}) for the combined system, the {\em black diamonds} are obtained by summing the optical depths of the two profiles computed using \scipy. Using the convolution of Eq.~(\ref{eq:damping}) yields very accurate Voigt profiles.} 
     \label{fig:dla}
\end{figure*}

\begin{figure*}
     \includegraphics[width=\textwidth]{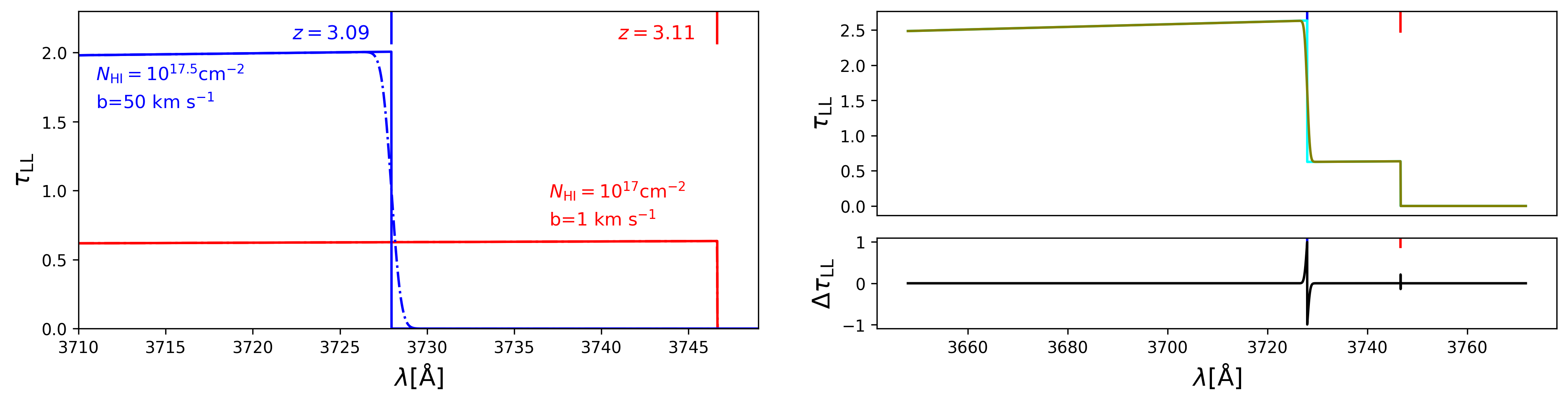}
      \caption{Illustration of the implementation of Lyman-limit absorption. The spectrum contains two \lls s with parameters $N_{\rm HI}=10^{17.0}~{\rm cm}^{-2}$, thermal broadening $b_D=1~{\rm km~s}^{-1}$ and redshift $z=3.11$
      (red absorber) and $N_{\rm HI}=10^{17.5}~{\rm cm}^{-2}$, thermal broadening $b_D=50~{\rm km~s}^{-1}$ and redshift $z=3.09$ (blue absorber).
      In the {\em left panel}, we plot the optical depth $\tau$
      due to Lyman limit absorption for each absorber separately, with the {\em dot-dashed line}
      our calculation using Eq.~(\ref{eq:limit}), and the {\em solid line} showing
      $\tau=\tau_\sigma=N_{\rm HI}\sigma(\lambda/(1+z))$, which is the Lyman-limit optical depth in the absence of thermal broadening. Since $b_D$ is small for the red absorber, solid and dot-dashed lines overlap, whereas they differ for the blue absorber. In the {\em right panel}, we plot the optical depth due to the two absorbers combined. The {\em green line} is the sum of the non-thermally broadened optical depths, and the {\em cyan line} is computed using Eq.~(\ref{eq:limit}). The {\em lower right panel} shows the difference between the cyan and green optical depths. Using the convolution of Eq.~(\ref{eq:limit}) yields very accurate ionisation edges.}
      \label{fig:lls}
\end{figure*}

\label{sect:damping}

\begin{table}[]
\centering
\caption{Values of the vacuum rest-wavelength ($\lambda_n$), oscillator strength ($f_n$), and line broadening parameter ($\Delta v_{\gamma, n}$) for selected Lyman-series lines $1\to n$ in the hydrogen atom, computed as described in Appendix \ref{sect:hydrogen_data}.}
\begin{tabular}{ccccl}
\hline
$n$       & $\lambda_n$ & $f_n$  & $\Delta v_{\gamma, n}$  \\ 
& [\AA] & [-] & [${\rm cm}~{\rm s}^{-1}$] \\
\hline\hline\\
2 & 1215.684 & $4.16197\times 10^{-1}$ &  $6.05752\times 10^{2}$ \\
 3 & 1025.734 & $7.91016\times 10^{-2}$ &  $1.54762\times 10^{2}$ \\
 4 & 972.547 & $2.89910\times 10^{-2}$ &  $6.28618\times 10^{1}$ \\
 5 & 949.753 & $1.39383\times 10^{-2}$ &  $3.17539\times 10^{1}$ \\
10 & 920.973 & $1.60537\times 10^{-3}$ &  $3.93031\times 10^0$ \\
20 & 914.048 & $1.96675\times 10^{-4}$ &  $4.91987\times 10^{-1}$ \\
30 & 912.777 & $5.80583\times 10^{-5}$ &  $1.45912\times 10^{-1}$ \\
40 & 912.333 & $2.44616\times 10^{-5}$ &  $6.15841\times 10^{-2}$ \\
60 & 912.017 & $7.24117\times 10^{-6}$ &  $1.82556\times 10^{-2}$ \\
80 & 911.906 & $3.05388\times 10^{-6}$ &  $7.70301\times 10^{-3}$ \\
100 & 911.854 & $1.56335\times 10^{-6}$ &  $3.94418\times 10^{-3}$ \\
\hline
\end{tabular}
\label{tab:lyman}
\end{table}


The line profile of Eq.~(\ref{eq:gauss}) assumes that the electronic transition itself is a Dirac delta function in wavelength. However, excited levels have a finite life time resulting in spontaneous decay as described by the Einstein $A$ coefficient. The finite life time leads to intrinsic line broadening\footnote{If the particle density is high, for example in a stellar atmosphere, collisions may limit the life time of levels as well which increases the line width by an amount that depends on density and temperature, see e.g. \citet{Yelnik81}. Densities in the \ism{} are sufficiently low that this effect is negligible.}. The intrinsic line shape in velocity space is often approximated by a Lorentzian (subscript D for \lq damping\rq)
\begin{align}
    \Phi_{\rm D}(v) &= \frac{1}{\pi}\,\frac{\Delta v_\gamma} {v^2+\Delta v_{\gamma}^2}\,;\quad \Delta v_{\gamma} = \frac{\lambda_0\,\gamma}{4\pi}\,.
    \label{eq:Lorentz}
\end{align}
Here, $\lambda_0$ is again the vacuum rest-wavelength of the transition and $\gamma$ is the inverse of the life time. Due to the analogy with a damped oscillator, $\gamma$ is called the damping constant, and systems in which the Lyman-$\alpha$ damping wing is prominent, which have $N_{\rm HI}\ge 10^{20.3}{\rm cm}^{-2}$, are called \lq damped Lyman-$\alpha$\rq\ systems (\dla{}s); see e.g. \citet[][]{Rybicki86} p.~289. The shape of the natural line profile is only approximately that of the Lorentzian of Eq.~(\ref{eq:Lorentz}), a more accurate line profile is discussed by \cite{Lee20}, who also provide historical references on how this profile came about.

Hydrogen's \lya\ transition has two components, which are ${^2P_{3/2}}\to {^2S_{1/2}}$ and ${^2P_{1/2}}\to {^2S_{1/2}}$ in the usual notation of energy-angular momentum levels. The Einstein coefficients for these levels are nearly identical, $6.2647\times 10^8$ and $6.2648\times 10^8~{\rm s}^{-1}$, as are the rest wavelengths, $1,215.668237310$ and $1,215.673644608$\AA, according to the Atomic Spectra Database Lines form\footnote{\href{https://physics.nist.gov/PhysRefData/ASD/lines_form.html}{https://physics.nist.gov/PhysRefData/ASD/lines\_form.html }}
(hereafter {\sc nist}). Currently, \spec{} accounts for hydrogen \lya\ absorption with a single component (i.e. ignoring this doublet structure), although it would be straightforward to improve on this. The line broadening of this (single) line is


\begin{align}
    \Delta v_\gamma(n=2) &= \frac{\lambda\,\gamma}{4\pi}
    \approx 606.05~{\rm cm}{\rm ~s}^{-1}\,,
\end{align}
where $\gamma=A_{2,1}$ for the \lya{} line.

Natural and Doppler broadening can be combined by convolving the Gaussian profile of Eq.~(\ref{eq:gauss}) with the Lorentzian profile of Eq.~(\ref{eq:Lorentz}), resulting in a Voigt profile in the case of a single absorber. Alternatively and more conveniently, the optical depth including the effects of the damping wing, $\tau_{\rm D}$, is the convolution of the optical depth without a damping wing, $\tau$, with the Lorentzian, $\Phi_{\rm D}$,
\begin{align}
    \tau_{\rm D}(v) = \int_{-\infty}^{\infty}\,\tau(v')\,\Phi_{\rm D}(v-v')\,dv'\,.
    \label{eq:damping}
\end{align}
We use convolution to account for the damping wing of the Lyman-series lines of \ion{H}{i} as described in \S \ref{sect:convolution} (see also \citealt{Wijers20}).

\subsubsection{Higher order Lyman series lines}
Values of the vacuum rest-wavelength, $\lambda_n$ and oscillator strength, $f_n$, for the $1\to n$ Lyman series transitions are known analytically for the hydrogen atom, see Appendix \ref{sect:hydrogen_data} for details.

The damping constant for these lines is the sum of the Einstein coefficients from the upper level, $n$, to all lower levels, $n'$,
\begin{align}
    \gamma_n = \sum_{n'=1}^{n-1}\,\sum_{l=0}^{n'-1} A_{n1, n'l'}\,.
    \label{eq:gamma}
\end{align}
(Using the usual notation where $n$ is the principal quantum number and $l$ the angular momentum quantum number). For example, Lyman-$\beta$ has $n=3$ and damping constant $\gamma_3=A_{31, 20}+A_{31, 10}$. If the lower-energy state itself has a finite life time, then this would also contribute to $\gamma_n$. Of course, the lower-energy state for Lyman-series lines is the ground state, which is stable. We compute the $A$'s using the {\sc fortran} programme described by \cite{hoang-binh_program_2005}, which can compute the $A$'s up to $n=1000$. A copy of this code is part of the \spec{} package. Given these $A$'s, we compute the damping constant using Eq.~(\ref{eq:gamma}).

We have compared our computed values of $\lambda_n$, $f_n$ and $\gamma_n$ to literature values. We find agreement with the values for $\lambda_n$ and $f_n$ in the file \texttt{atom.dat}, which is part of the line-fitting programme {\sc vpfit} of \cite{Carswell14}, to better than 0.1~per cent up to $n=6$, but at higher\footnote{\texttt{atom.dat} contains data for the Lyman series up to $n=31$.} $n$, our values are $\sim 30$ per cent larger. Our values agree to better than 0.01~per cent with those quoted in the {\sc nist} database, which contains values up to $n=40$. We report the values of $\lambda_n$, $f_n$ and damping constant $\Delta v_{\gamma,n}$ for some illustrative values of $n$ in Table~\ref{tab:lyman}. We implement higher-order lines using convolution, as described in \S \ref{sect:convolution}.





\subsubsection{Bound-free transitions: continuum absorption}
\label{sect:limit}
Photons with energy $E=h\nu\ge E_{\rm th}$ can ionize \ion{H}{i} and this introduces an \lq ionisation edge\rq\ and continuum absorption above it - as opposed to a line - in the spectrum; $E_{\rm th}$ is the ionisation energy. Systems for which this edge is very prominent in spectra have $N_{\rm HI}\ge 10^{17.2}{\rm cm}^{-2}$ and are called Lyman limit systems (\lls{}s; the \lq limit\rq\ refers to the fact that the edge occurs at the energy of Lyman lines in the limit of $n\to\infty$). The photoionisation cross section $\sigma_{\rm LL}$ is energy dependent, and we evaluate it using the fitting function proposed by \cite{Vern96sigmacross}, reproduced in Appendix~\ref{sect:sigma}.

The continuum optical depth at Doppler velocity $v$ due to \ion{H}{i} ions with Doppler velocity $v'$, is
\begin{align}
    d\tau_{\rm LL}(v, v') &= \sigma_{\rm LL}\left(E(v-v')\right)\,\frac{dN_{\rm HI}(v')}{dv'}\,dv'\,,
\end{align}
where the photon energy $E(v-v')=\exp(-(v-v')/c)~E_{\rm th}$. As before, we stress that the use of velocity rather than energy is simply a change of variables rather than a Doppler boost: there are no relativistic corrections (except for such high energies that the cross-section fit $\sigma_{\rm LL}(E)$ itself requires such corrections). We now eliminate $dN_{\rm HI}(v')/dv'$ in terms of $\tau(v')$ using Eq.~(\ref{eq:tgauss}). Integrating over $v'$ yields the continuum optical depth in terms of $\tau$,
\begin{align}
    \tau_{\rm LL}(v) &= \int_{-\infty}^{\infty}\,\tau(v')\frac{\sigma_{\rm LL}\left(E(v-v')\right)}{\sigma}\,\frac{dv'}{c}\,\,.
    \label{eq:limit}
\end{align}
The observed wavelength that corresponds to velocity $v$ is now $\lambda(v)=\lambda_{\rm LL}\,(1+z)\exp(v/c)$, where $\lambda_{\rm LL}=hc/E_{\rm th}\approx 911.27$\AA\ is the vacuum wavelength of a photon with energy $E_{\rm th}$. We use convolution to account for the ionisation edge of the \ion{H}{i} series as well, as described in more detail next.

\subsubsection{Implementation of the Lyman series, damping wings and continuum absorption using convolution}
\label{sect:convolution}
Higher-order \lya{} lines can be accounted for using convolution because the line-width ($b_D$-parameter) is the same for all of them. Therefore, the optical depth due to higher-order Lyman lines can be derived by convolving the \lya-optical depth. For the damping wing and continuum absorption, convolution is possible because the damping and continuum absorption parameters are physical constants and hence are also the same for all lines.


To account for Lyman-series lines, we define the following convolution kernel,
\begin{align}
    K_{\rm HI}(v) &= \sum_{n=3}^{N_{\rm Lyman}+1} {\cal H}_n\,\Phi(v-v_2-v_n)\,,
    \label{eq:HHI}
\end{align}
where
\begin{align}
    {\cal H}_n &= \frac{f_n\,\lambda_n}{f_{2}\,\lambda_{2}}\,;\quad
    v_n = c\,\ln(\frac{\lambda_{n}}{\lambda_{2}})\,.
    \label{eq:H}
\end{align}
Here, $N_{\rm Lyman}$ is the number of Lyman series we want to account for and $\Phi$ is the line-shape of the \lya{} line. The offset $v_2+v_n$ in the line shape accounts for the difference in wavelength between the particular Lyman line and \lya\ and the scaling factor ${\cal H}_n$ accounts for the difference in 
cross section.

For those lines for which we want to include a damping wing, we take $\Phi$ to be the Lorentzian of Eq.~(\ref{eq:Lorentz}), where the damping constant $\gamma_n$ is computed from Eq.~(\ref{eq:gamma}). The values of the damping constant, $\gamma_n$, decrease rapidly with increasing $n$, approximately $\propto n^{-3}$. Therefore it is a good approximation for high enough $n$ to replace the Lorentzian by a Dirac delta function,

\begin{align}
\Phi(v) &\to \delta^{\rm Dirac} \nonumber\\
&= 
         \begin{cases}
(\Delta v_{\rm pix})^{-1} & \text{if $v=0$}\\
            0, & \text{otherwise.}
		 \end{cases}
\label{eq:dirac}
\end{align}
The expression on the second line is our numerical implementation of the Dirac delta function, where $\Delta v_{\rm pix}$ is the (constant) pixel size used in the generation of spectra.

Approaching the ionisation threshold, at $\lambda\approx  911.763$\AA,  the Lyman series lines get closer together and simultaneously get weaker. Line separation and line strength conspire to make the optical depth independent of wavelength sufficiently close to the threshold. To see how this comes about, we compute the convolution kernel averaged over a narrow velocity range close to - but to the red of - the threshold,
\begin{align}
    {\bar K}_{\rm HI}(v) &= \lim_{\delta v\to 0}\frac{\int_{v_n}^{v_n+\delta v} K_{\rm HI} (v)\,dv}{\delta v}\nonumber\\
    &=\lim_{n\to\infty} \frac{(f_n\lambda_n)/(f_2\lambda_2)}{c\,\Delta \lambda/\lambda_n}\nonumber\\
    &=\lim_{n\to\infty} \frac{(f_n\lambda_n)/(f_2\lambda_2)}{2c/n^3}\,.
\end{align}
For large values of $n$, $f_n\to 256/(3e^4n^3)$, hence
\begin{align}
    {\bar K}_{\rm HI}(v) &\to \frac{128}{3e^4\,c}\,\frac{R^{-1}}{f_2\,\lambda_2}\,
\end{align}
where $R$ is the Rydberg constant of \ion{H}{i} from Eq.~(\ref{eq:R}) below.
Since the bunching-up of Lyman-series lines on the red side of the Lyman limit makes $K_{\rm HI}$ approximately constant, we can perform the convolution integral in Eq.~(\ref{eq:fullconv}) trivially, to find that the optical depth close to the Lyman limit is independent of wavelength and thermal or turbulent broadening, and is given by
\begin{align}
\lim_{n\to \infty}\tau({\rm Ly}_n)
&= \frac{128}{3e^4}\,\left(\frac{3\pi\sigma_T}{8}\right)^{1/2}\,R^{-1} N_{\rm HI}\nonumber\\
&\approx \frac{N_{\rm HI}}{10^{17.200}\,{\rm cm}^{-2}}\,.
\end{align}
Not surprisingly, this equals the value of the cross section at the Lyman edge: Lyman lines with high $n$ generate an optical depth that is independent of wavelength and which smoothly joins the value of the cross section at the ionisation threshold.

The kernel that accounts for \ion{H}{i}'s bound-free transitions is
\begin{align}
  K_{\rm I}(v) &= \frac{\Sigma(v-v_2-v_{\rm LL})}{\sigma\,c}\,.
\end{align}
Here, $\sigma$ is the cross section of the \lya\ transition from Eq.~(\ref{eq:sigma}), $v_2$ the velocity from Eq.~(\ref{eq:H}) for $n=2$, and $\Sigma$ the bound-free cross section from Eq.~(\ref{eq:Sigma}) transformed from photon energy to velocity offset.

Finally, we perform the following convolution,
\begin{align}
    \tau(v) &= \int_{-\infty}^{\infty} \tau_{\rm G}(v')\,K(v-v')\,dv'\nonumber\\
    K(v) &= K_{\rm HI}(v) + K_{\rm I}(v)\,,
\label{eq:fullconv}    
\end{align}
where $\tau_{\rm G}$ is the \lya-optical depth in the Gaussian approximation. This convolution accounts for ({\em i}) all Lyman lines up to $N_{\rm Lyman}$, ({\em ii}) damping wings for all Lyman lines up to that value of $n$ where the Lorentzian line profile function is used, and ({\em iii}) thermal and turbulent broadening - which is the same for all Lyman lines.

\subsection{Tests of the Lyman-series implementation}
We use the \scipy\ \python\ function {\sc convolve} to perform the convolutions discussed in the previous section. The velocity grid over which the convolution is applied needs to be high resolution, of order $\Delta v_{\rm pix}\approx 10^{-3}\unit{km~s^{-1}}$. This is because the Lorentzian profile has a sharp peak at zero lag {\em and} extended wings. The shape of the bound-free cross-section is much smoother and does not require such a high-resolution velocity grid. For that reason, it is computationally more efficient to perform the two convolutions (for the Lyman series and their damping wings, versus that for the continuum absorption) separately.

We illustrate this calculation in Fig.~\ref{fig:edge}, where we plot the transmission $\exp(-\tau)$ of a single absorber with column density $10^{17.3}{\rm cm}^{-2}$ and thermal and turbulent broadening parameters of $12~{\rm km~s}^{-1}$, in the rest-frame of the absorber. The spectrum is computed as described above, accounting for 40,000 Lyman series lines, 40 of which include a damping wing. Orange lines show individual Lyman series lines up to $n=100$; for higher $n$, only every 100$^{\rm th}$ line is shown (vertical ticks indicate the wavelength of the line, the approximately Gaussian profile is the line's transmission, $\exp(-\tau)$). The sum of all these lines is shown in red.  Bound-free absorption is shown by the blue line. This absorption extends to the red side of the ionisation threshold - and, similarly, line absorption extends to the blue side of the threshold - due to the assumed thermal and turbulent broadening of the absorber. The level of absorption due to bound-bound transitions (lines) and bound-free transitions is continuous through the ionisation threshold, as shown by the purple line, for which the optical depth is the sum of the line and bound-free optical depths. The previous section explains the reason 
why.

We demonstrate the accuracy of the damping wing implementation by initialising a sight line where the \ion{H}{i} column density in each pixel is zero everywhere, except in two pixels where it is high enough that the damping wing should be prominent. Specifying the temperature of the \ion{H}{i} gas in those two pixels, we use the \scipy\ implementation of the Voigt profile (function {\sc voigt\_profile})  to compute the optical depth. We compare the result to our convolution implementation in Figure~\ref{fig:dla}. The top panel of that figure shows $\exp(-\tau)$, the lower panel shows the corresponding optical depth, $\tau$. The small inset shows that the convolution integration agrees very well with the line shape computed by the \scipy\ function. The optical depth becomes very large close to the line centre, $\tau\approx 10^6$: there, the two methods differ in the value $\tau$ by order unity.

We perform a similar test for the \ion{H}{i} ionisation edge in Figure~\ref{fig:lls}. The red absorber has a negligible thermal width of $b_D=1~{\rm km~s}^{-1}$, and hence the optical depth profile is approximately set by the shape of the photo-ionisation cross section. The calculation using convolution agrees very well with a direct implementation of the cross section, $\sigma_{\rm LL}\left(E(v-v')\right)$ (right panel). The blue absorber has a non-negligible thermal width of $b_D=50~{\rm km~s}^{-1}$ and this larger thermal broadening smears out the profile at the absorption edge.

We note that the advantage of the convolution method is that there is no need to identify \dla s nor \lls s in the spectrum: simply convolving the optical depth in the Gaussian approximation automatically accounts for the natural line profile of any Lyman-series line (independent of column density), and similarly, any absorber will generate its proper share to the optical depth due to bound-free absorption.

\subsection{Atomic data for other ions}
\label{sect:atomic}
\begin{table}[]
\centering
\caption{Sample output of the \lq A-Z\rq\ of atomic constants collected from the \nist\ database. For all resonant transitions of all ions of all elements of the periodic table, we collect the laboratory wavelength ($\lambda_0$) and oscillator strength (\lq $f$\rq-value) and store them in an {\sc hdf5} table. The table contains a total of 20, 226 lines from 335 ions of 60 elements.}
\begin{tabular}{ccccl}
\hline
Element   & Ion      & $\lambda_0$  & f-value  &  \\ 
          &          & (\AA)        &          & \\    \hline
Actinium  &          &                                  &          &  \\
          & Ac I     &                                  &          &  \\
          &          & 4181.1552                        & 0.324    &  \\
          &          & 4181.1552                        & 0.391    &  \\
          &          & $\vdots$                         & $\vdots$ &  \\
          & $\vdots$ &                                  &          &  \\
$\vdots$  &          &                                  &          &  \\
Hydrogen  &          &                                  &          &  \\
          & H I      &                                  &          &  \\
          &          & 1215.6701                        & 0.41641  &  \\
          &          & 1025.722                         & 0.079142 &  \\
          &          & 972.5365                         & 0.029006 &  \\
          &          & $\vdots$                         & $\vdots$ &  \\
$\vdots$  &          &                                  &          &  \\
Zirconium &          &                                  &          &  \\
          & $\vdots$ &                                  &          &  \\
          & Zr VI    &                                  &          &  \\
          &          & 568.285                          & 0.029    &  \\
          &          & $\vdots$                         & $\vdots$ & \\
\hline
\end{tabular}

\label{table:fl}
\end{table}

The list of transitions included when modelling a quasar spectrum - either when generating a mock spectrum or fitting an observed spectrum - 
emerged historically and pragmatically. The choice was also influenced by
the paper by \cite{Morton91}, who computed oscillator strengths and rest wavelengths for specific transitions.  In \spec{} we instead extract {\em all} resonant transitions 
that appear in \nist. The \python\ class that generates the query is part of the \spec\ distribution.
The query returns 20, 226 resonant transitions from 335 ions of 60 elements, and we store the laboratory wavelength, $\lambda_0$, oscillator strength, $f$, and the terms that describe the upper and lower level in an {\sc hdf5} file. 

We further use the \textsc{mendeleev} python package \footnote{\url{https://mendeleev.readthedocs.io/en/stable/}}
\citep{mendeleev2014} to obtain the atomic masses of these elements.
Table~\ref{table:fl} illustrates this \lq A to Z\rq\ of atomic constants contained in the {\sc hdf5} table by listing atomic constants from Actinium to Zirconium. The file also contains
40 \ion{H}{i} Lyman-series lines. We do not use these parameters but rather compute them for hydrogen (and deuterium) as described above. 

The \spec{} user can specify which ions they want included in the calculation. As described before, \spec\ will compute the optical depth for the transition with the largest oscillator strength for that ion in the form of short spectra.
The optical depth for any other line of that ion is a simple scaling of the optical depth of the strongest line. For example, for the doublet at $\lambda=1550, 1538$\AA~ of \ion{C}{iv}, we compute the strongest component from the particle projection, and then obtain the weaker component by scaling the optical depth by the ratio of $\lambda_0\,f$ of the two. We note that such scaling is extremely fast: the \spec{} run time is mostly set by how many ions are included in the calculation and whether damping wings/ionisation edges are included or not: the actual number of lines used has little impact on compute time. For reference, a calculation of a single spectrum through an \eagle\ L0100N1504 simulation (1504$^3$ particles) using 3 ions takes 24\unit{s} to read the data, 9\unit{s} to project the data, and 10\unit{s} to compute the optical depths and other spectral properties. Data reading is relatively fast, because the snapshot uses a spatial hash-table to identify which particles intersect the chosen sight line. We discuss examples that use this simulation next.

\ifSubfilesClassLoaded{\bibliography{mybib}}
\end{document}

\section{Example applications of \spec{}}
We have chosen representative examples of mock spectra to illustrate the use and versatility of \spec. Most examples are based on data from the {\sc eagle} project \citep{Schaye15, Crain15}, whose public data are described by \cite{McAlpine16}, see also \cite{Eagle17}. In particular, we use data from the simulation labelled L0100N1504 in table~1 of \cite{Schaye15}. Briefly, this is an {\sc sph} simulation performed using the {\sc gadget-3} incarnation of {\sc gadget} \citep{Springel05}, with changes to the {\sc sph} scheme and added functionality for sub-grid recipes of galaxy formation, namely, star formation and the thermal and chemical feedback from stars, as well as black hole seeding and growth, and feedback from accreting black holes. The simulation follows the production, destruction, and release of 11 elements of the periodic table (namely H, He, C, N, O, Ne, Mg, Si, S, Ca and Fe), as described by \cite{Wiersma09b}. \spec\ uses the abundances of these elements to compute any absorption by their ions. The cubic simulation volume is $L=$100 co-moving megaparsecs (cMpc) on a side, and the gas and dark matter are represented with $1504^3$ particles of each type. The gas is photo-heated and cools radiatively in the presence of an
ambient (optically thin) \cite{Haar12UVbackground} ionising background, with radiative cooling tracked element-by-element following \cite{Wiersma09}. The mass of a gas particle is $\approx 2\times 10^6{\rm M}_\odot$. We also use \eagle\ simulation {\sc recal}L0050N1504 ($L=50$~Mpc, 8-times better mass resolution, which uses the \lq {\sc recal}\rq\ choices of parameters from \cite{Schaye15}). This simulation was not part of the original \eagle\ project.

We use these simulations together with \spec{} to illustrate the calculation of \ion{H}{i} \lya{} forest spectra and discuss their statistics. We study correlations between \ion{H}{i} and metal lines, and we investigate the statistics of metal lines near galaxies.  In all examples, we generate spectra with pixel size of 1\unit{km~s^{-1}} which is better than that of the typical high-resolution echelle spectrograph. The resolution of the mock spectra is therefore limited by that of the simulation - rather than by \spec.

\subsection{\lya\ forest examples}
\label{sect:Lya}
At each redshift, we shoot $250$ sight lines along a Cartesian axis through the corresponding L0100N1504 \eagle\ snapshot, and use \spec{}  to compute the \ion{H}{i} \lya\ optical depth along them. We then decompose each spectrum in \lq absorption lines\rq, using the decomposition method introduced by \cite{Garzilli15}. The decomposition identifies a line as a stretch of spectrum between two successive local minima in the optical depth. This type of decomposition is not straightforward in an observed spectrum due to noise, and the lines identified using decomposition are not identical to those identified using a line-fitting method such as {\sc vpfit}. However, we think that line statistics are sufficiently similar to warrant this analysis.

The \ion{H}{i} column density of the absorber producing a line is computed by integrating Eq.~(\ref{eq:tgauss}),
\begin{align}
    N_{\rm HI} &= \frac{1}{\sigma\,c}\,\int_{v_{\rm min}}^{v_{\rm max}}\,\tau(v)\,dv\,,
\end{align}
where $v_{\rm min}$ and $v_{\rm max}$ are the locations of the local minima that define the line; as before, $\sigma$ is the cross section of the transition. We use this decomposition to determine the fraction of light absorbed due to lines in a given column density range. We perform this analysis at a range of redshifts.
We use this line decomposition to illustrate how lines of various column density affect \lya{} forest statistics. One of these statistics is the mean transmission, which we discuss next.

\subsubsection{The evolution of the mean transmission}

\begin{figure}
     \includegraphics[width=\columnwidth]{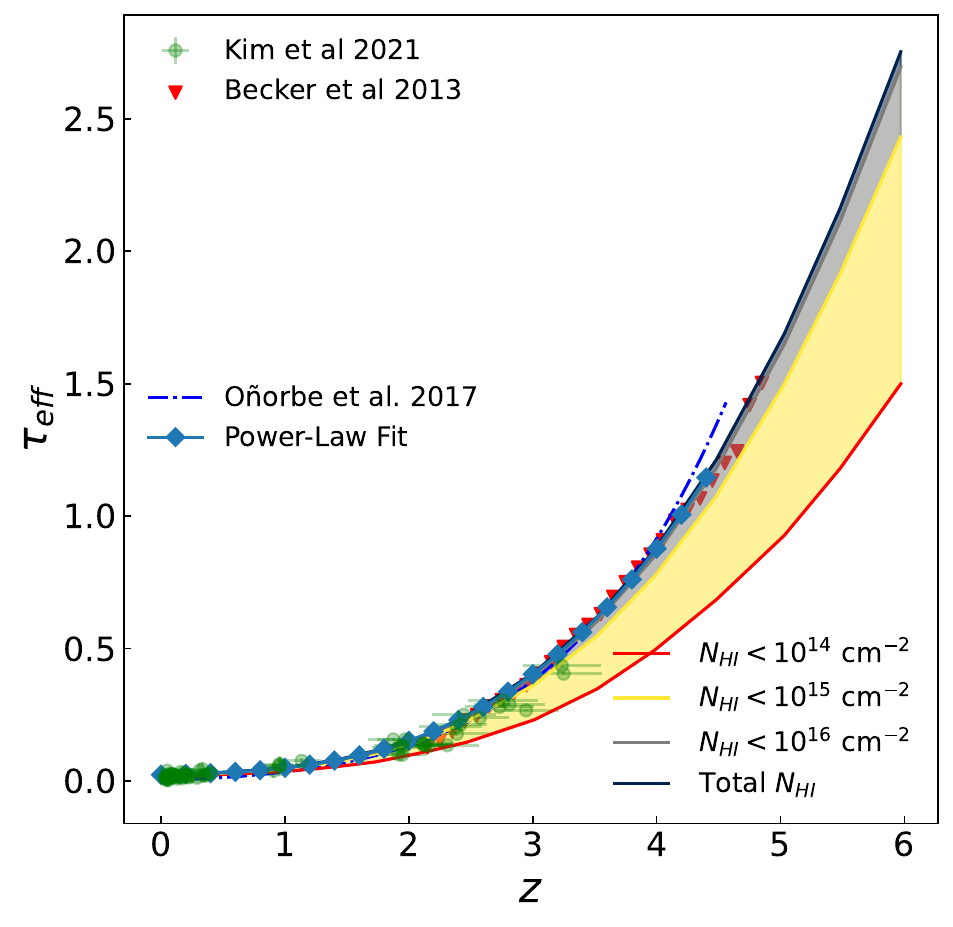}
      \caption{Evolution of the \lya\ effective optical depth, $\tau_{\rm eff}$.
      Observational data from \protect\cite{Kim21} are shown as {\em green dots} and of \protect\cite{Becker2013} as {\em red triangles}.
      The polynomial fit of Eq.~(\ref{eq:poly_fit}) is shown as a {\em blue line connecting diamonds}, and the fit of \protect\cite{Onorbeb17} as a {\em blue dash-dotted line}. The simulated evolution, calculated using \spec\ for the \eagle\ simulation, is shown as the {\em dark blue line}. It is scaled to reproduce the polynomial fit. The {\em red}, {\em yellow} and {\em grey} line show $\tau_{\rm eff}$ when including lines up to
      $\log N_{\rm HI}[{\rm cm}^{-2}]$ of 14, 15 and 16: this shows that most of the absorption is due to absorbers with $\log N_{\rm HI}[{\rm cm}^{-2}]< 15$.}
\label{fig:fluxevol}
\end{figure}

\begin{figure}
     \includegraphics[width=\columnwidth]{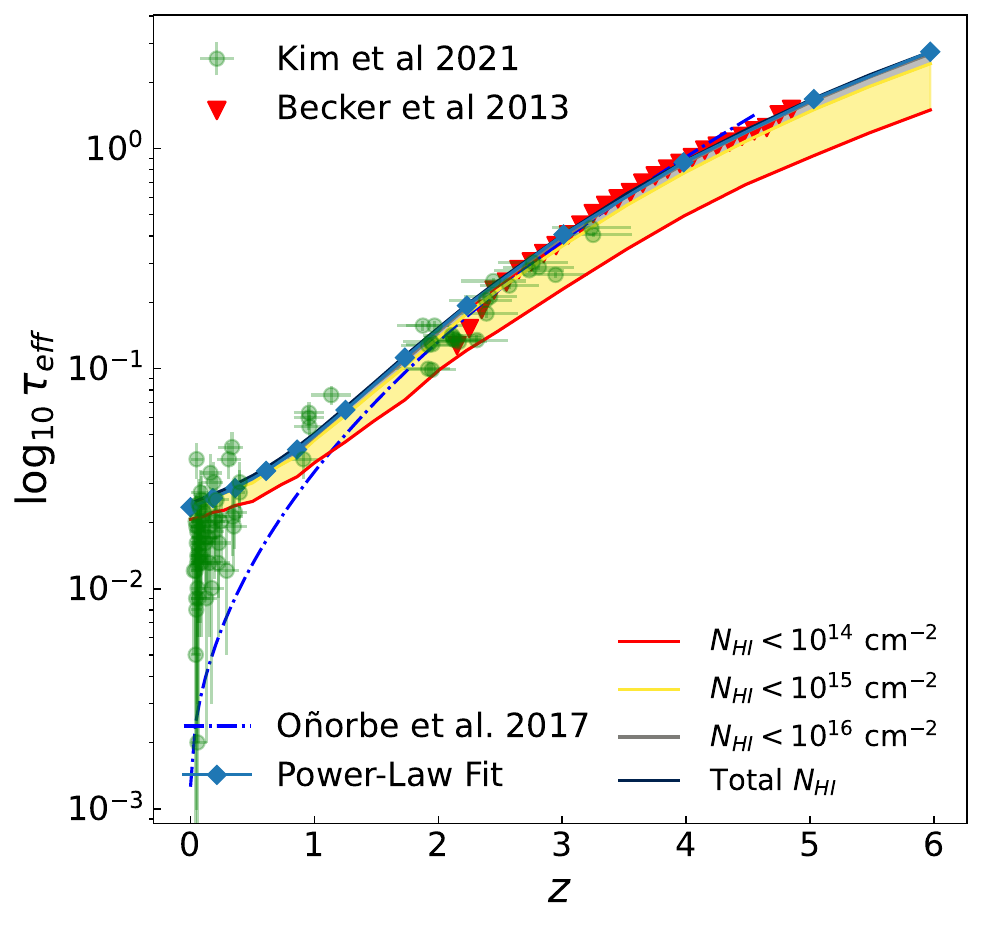}
      \caption{Same as Fig.~\ref{fig:fluxevol}, but plotting $\log\tau_{\rm eff}$, to better show the large scatter in the observed value of $\tau_{\rm eff}$ at $z\le 1$, which results in a relatively large relative uncertainty in the evolution of $\tau_{\rm eff}$. We also note that the high-resolution data ({\em green filled circles}) tend to have lower values of $\tau_{\rm eff}$ than the lower-resolution data ({\em red triangles}) where they overlap in redshift. 
      }
\label{fig:teff_log}
\end{figure}

\begin{figure}
     \includegraphics[width=\columnwidth]{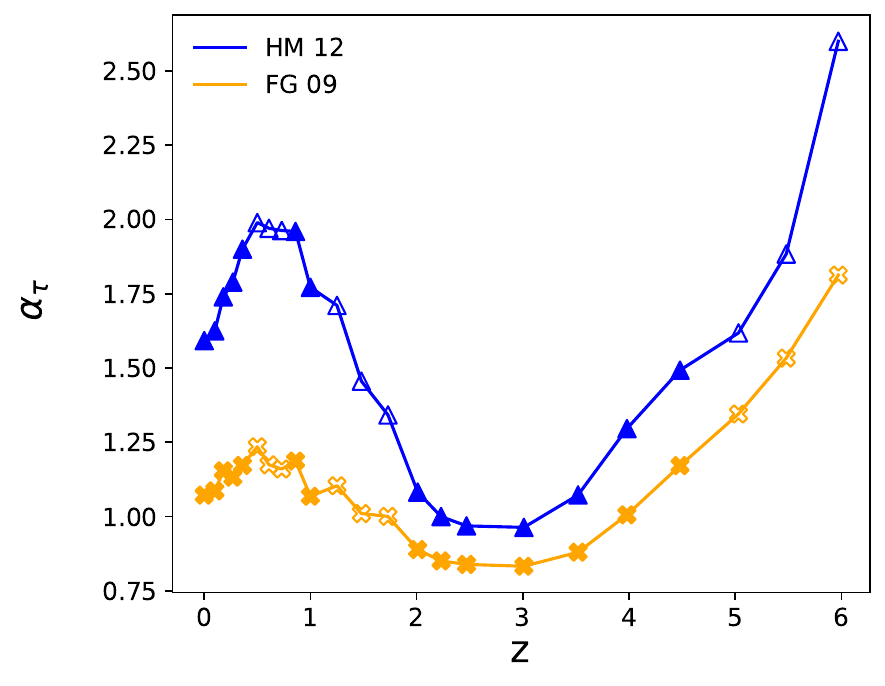}
      \caption{Required scaling factor, $\alpha_\tau$, for multiplying the \ion{H}{i} optical depth to reproduce the polynomial fit to the evolution of $\tau_{\rm eff}$ from Fig.~\ref{fig:fluxevol}. {\em Solid symbols} are less than $\Delta z = 0.1$ from observed values of $\tau_{\rm eff}$, {\em open symbols} are  extrapolated using the polynomial fit Eq.~(\ref{eq:poly_fit}). $\alpha_\tau$ is plotted in {\em blue} and {\em orange} for the \protect\cite{Haar12UVbackground} and \protect\cite{Fauc09} evolution of the ionising background.} 
\label{fig:tau_scaling}
\end{figure}

The mean transmission of a set of spectra is given by
\begin{align}
    \langle F\rangle &=\langle\exp(-\tau)\rangle\equiv \exp(-\tau_{\rm eff})\,,
\end{align}
where the angular brackets $\langle\cdots\rangle$ denote the ensemble average and $\tau_{\rm eff}$ is the effective optical depth. $\tau_{\rm eff}$ is a fundamental statistic that can be computed for both mock and observed spectra. The effective optical depth evolves with redshift, and depends primarily on the amplitude of the ionising background, $\Gamma(z)$, and in addition on the temperature-density distribution of the intervening gas. Because $\Gamma$ is not well constrained, simulators often scale $\Gamma$ until the simulated spectra have the same value of $\tau_{\rm eff}$ as the observations to which the simulations are compared. This scaling is possible because the temperature of the absorbing gas - which affects the properties of the lines as well - is independent of $\Gamma$ provided the gas is highly ionized. Observational determinations of $\tau_{\rm eff}$ \cite[e.g.][]{Schaye03, Becker2013, Kim21} are not completely straightforward because of uncertainties in continuum fitting and because some of the absorption may be due to ions other than hydrogen. Sample variance may also play a role \citep{Rollinde13}.

We plot $\tau_{\rm eff}(z)$ in Fig.~\ref{fig:fluxevol} and $\log(\tau_{\rm eff})$ in Fig.~\ref{fig:teff_log}, with observational determinations taken from \cite{Kim21} (for $z\le 3.5$, green circles) and \cite{Becker2013} (for $z\in [2,5]$, red triangles). We also overplot a fit of the form
\begin{align}
   \tau_{\rm eff}(z) &=  A_0(1+z) + B_0(1+z)^2+ C_0(1+z)^3\,,
   \label{eq:poly_fit}
\end{align}        
(blue line connecting diamonds labelled \lq power-law fit\rq) with parameters $ A_0 = 0.0466$, $B_0 = -0.0355$ and $C_0 = 0.0122$. The fitting parameters were determined by combining the evolution of $\tau_{\rm eff}$ with the evolution of the flux {\sc pdf} and flux power spectrum, both of which are discussed below. The fit of \cite{Onorbeb17} is shown as well.

Decomposing the simulated spectra in lines, we compute the mean transmission due to lines with column density lower than $\log N_{\rm HI}[{\rm cm}^{-2}]=$ 14, 15, 16 and $\infty$ (red, yellow, grey, and black line, respectively). At any shown redshift, lines with column density $\log N_{\rm HI}[{\rm cm}^{-2}]>15$ contribute little to $\tau_{\rm eff}$. The figure also shows that lower column density lines contribute proportionally more to $\tau_{\rm eff}$ at lower redshifts.

We scaled the \ion{H}{i} optical depths from the simulations by a factor $\alpha_\tau(z)$, in order to reproduce the fit to $\tau_{\rm eff}$ from Eq.~(\ref{eq:poly_fit}). (This corresponds to replacing the photoionisation rate $\Gamma\to \Gamma/\alpha_\tau$.) The scaling factor $\alpha_\tau$ for scaling the \cite{Haar12UVbackground} and \cite{Fauc09} ionising backgrounds are plotted in Fig.~\ref{fig:tau_scaling}. The \cite{Haar12UVbackground} scaling factor is $\approx 1$ for redshifts between approximately 2 and 4, meaning that the model value of the ionising background reproduces the mean transmission. However, we find that $\alpha_\tau>1$ both at higher and lower redshift, implying that the amplitude of this background may be too high (by a factor of 2 or more above $z=5$, as well as for $z\sim 1$). The \cite{Fauc09} background reproduces $\tau_{\rm eff}$ to better than 25 per cent up to $z\sim 5$, but at even higher $z$ its amplitude may be a bit too high as well.


\subsubsection{The evolution of the transmission probability distribution function}
\begin{figure*}
 \includegraphics[width=\textwidth]{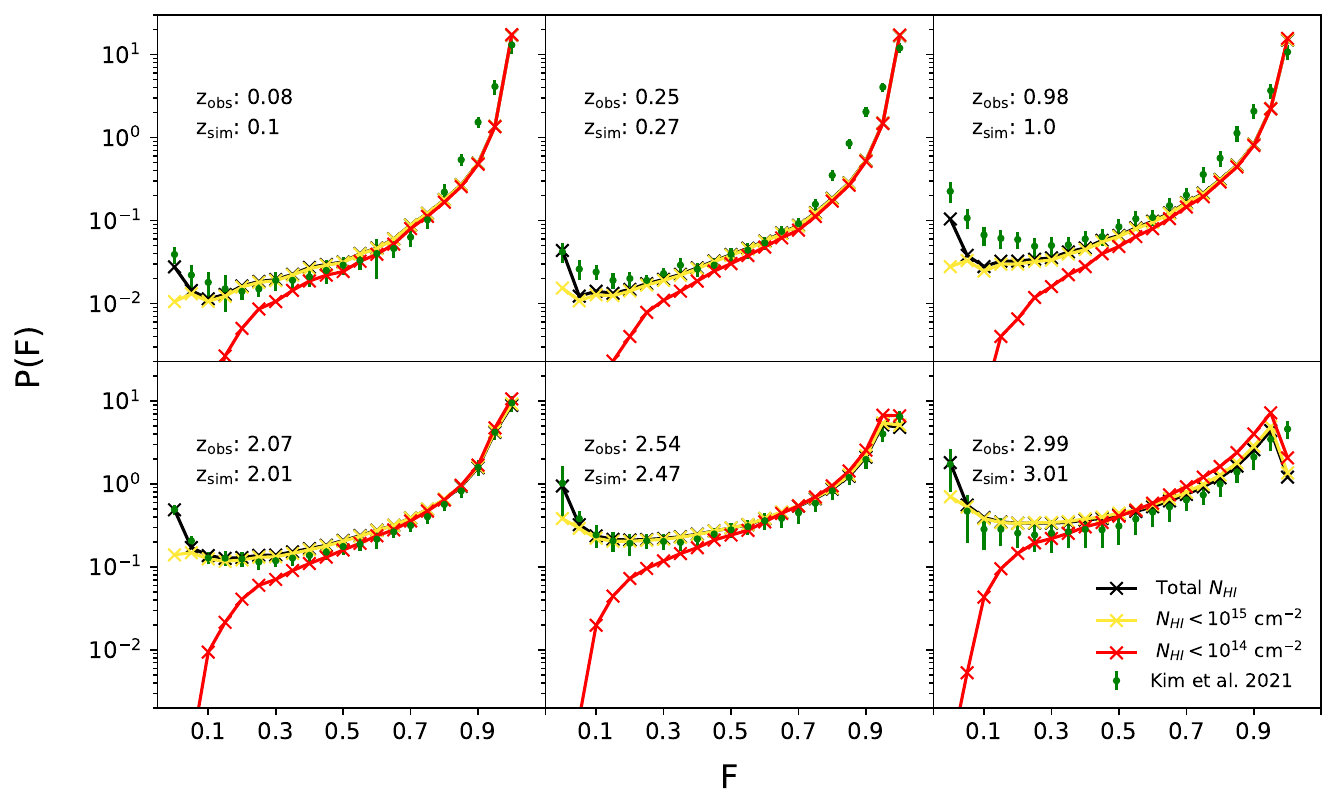}
\caption{Probability distribution function ({\sc pdf}) of the transmission in individual pixels, $F=\exp(-\tau)$, with different panels showing different redshifts. {\em Red symbols with error bars} are observations from \protect\cite{Kim21}. The {\em blue curves} are the results from the \eagle\ simulation scaled to
the evolution of the mean transmission from Eq.~(\ref{eq:poly_fit}). 
Each panel is labelled with the mean observed redshift, $z_{\rm obs}$, and the redshift of the simulation snapshot nearest to it, $z_{\rm sim}$. The {\em orange} and {\em green curves} show the normalised simulation {\sc pdf}'s when restricting the simulated spectra to lines with $\log N_{\rm HI}[{\rm cm}^{-2}]< 15$ and 14, respectively. The {\sc pdf} for $F\ge 1/2$ is dominated by weaker lines with $\log N_{\rm HI}[{\rm cm}^{-2}]< 14$, whereas for $F\le 0.1$, the stronger lines with $\log N_{\rm HI}[{\rm cm}^{-2}]\ge 16$ dominate. The simulated and observed histograms agree very well for all $z\ge 2$; see text for more discussion.}
\label{fig:FPDF}
\end{figure*}

\begin{figure}
 \includegraphics[width=\columnwidth]{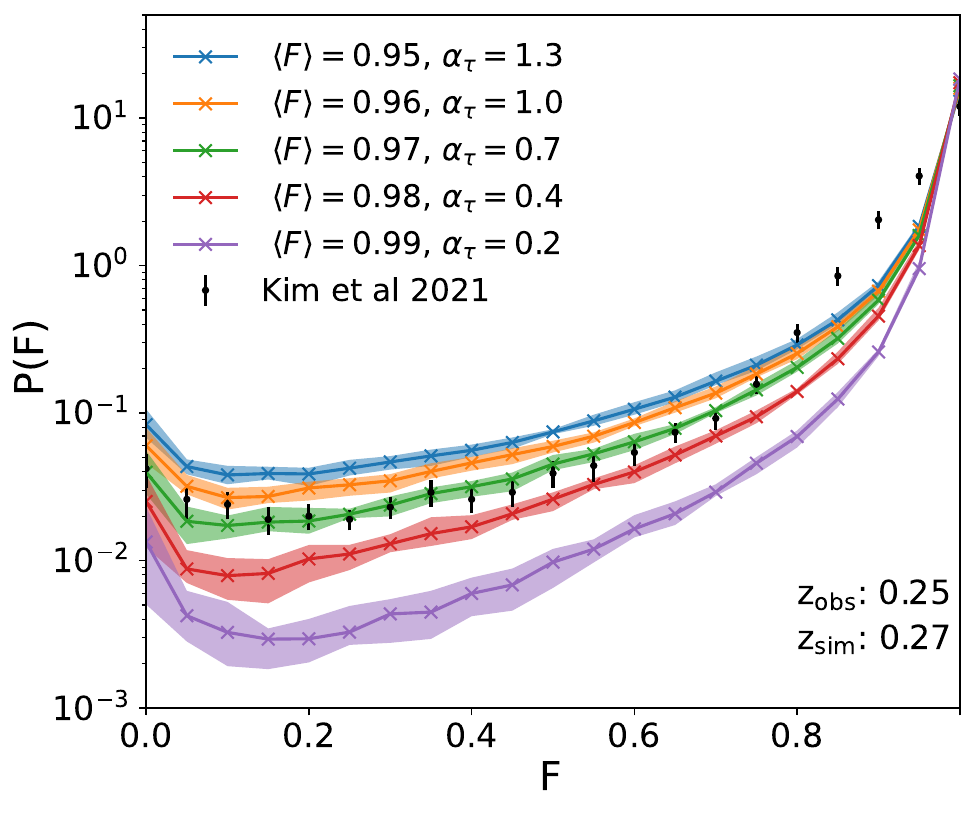}
\caption{Transmission {\sc pdf} at a $z\approx 0.25$, illustrating the dependence of the {\sc pdf} on the amplitude of the ionising background, $\Gamma$. {\em Black symbols with error bars} are observations at a mean observed redshift of $z_{\rm obs}=0.25$ from \protect\cite{Kim21}, reproduced from Fig.~\ref{fig:FPDF}. {\em Coloured lines} are for \eagle\ using the snapshot at redshift $z_{\rm sim}=0.27$, with coloured bands representing the variance of simulated spectra for realisations with the same absorption distance as the data. Different colours illustrate the impact of scaling the 
amplitude of the ionising background in the simulation, $\Gamma\to\Gamma/\alpha_\tau$, for different values of $\alpha_\tau$. The simulation curves are labelled by the value of $\alpha_\tau$; $\alpha_\tau=1$ corresponds to the amplitude of the \protect\cite{Haar12UVbackground} background.}
\label{fig:linebands}
\end{figure}

The transmission probability distribution function (usually called the \lq flux probability distribution\rq), $P(F)$, is a histogram of $F=\exp(-\tau)$ values, normalized to unit integral. $P(F)$ is plotted in Fig.~\ref{fig:FPDF} for a range of redshifts, with the data from \cite{Kim21} shown as red markers with error bars. The blue lines are $P(F)$ computed from \eagle\ (L0100N1504) using the scaling factor $\alpha_\tau$ that reproduces the evolution of the effective optical depth as shown in Fig.~\ref{fig:fluxevol}.

There is excellent agreement between simulation and observation for $z\ge\ 2$. We note in particular that \eagle\ reproduces the small increase in $P(F)$ around $F\approx 0$. This uptick is caused by strong lines with $\log N_{\rm HI}[{\rm cm}^{-2}]\ge 15$, as can be seen by comparing the blue and orange lines (the orange line is $P(F)$ for absorption lines with $\log N_{\rm HI}[{\rm cm}^{-2}]<15$, the blue line includes all absorption lines). The simulation also reproduces the shape of $P(F)$ close to the continuum, $F\approx 1$, except possibly for the highest $F\approx 1$ point at $z\approx 3$. The shape of $P(F)$ for $F\ge 0.7$ in the data depends rather sensitively on the placement of the continuum \cite[see e.g.][]{Rollinde13, Kim21}.

\cite{Bolton17} using the Sherwood suite of cosmological hydrodynamical simulations, compare their simulated $P(F)$ of a suite of simulations to high-resolution Echelle data from \cite{Kim07, Rollinde13} at $z=2.5$ and \cite{McDonald00} at $z=3.9$. Their simulations without feedback do not reproduce the increase in $P(F)$ at $F\approx 0$. Their simulations with feedback do slightly better, but still fail to reproduce the observed number of pixels with $F\approx 0$. We repeat that the \eagle\ simulation used here includes feedback (from stars and accreting black holes). \cite{Bolton17} also investigate in some detail the impact of uncertainties in continuum fitting on the shape of $P(F)$ for $F\ge 0.7$. Including a model for continuum-fitting errors improves the agreement with data, but differences between their simulations and the data persist.

The agreement between the mock spectra of \eagle\ and the observations is slightly worse for $z_{\rm obs}\le 1$ (top row of Fig.~\ref{fig:FPDF}), in particular for $F\gtrapprox 0.7$. \cite{Kim21} state that the shape of the observed $P(F)$ is sensitive to continuum placement and signal-to-noise, and restrict their analysis of $P(F)$ to the range $0.1\le F\le 0.7$. We find good agreement between \eagle\ and the observations for $F\le 0.7$ for $z_{\rm obs}=0.08$ and $0.25$, with poorer agreement for $z_{\rm obs}=0.98$.

The shape of $P(F)$ depends sensitively on $\Gamma(z)$, as illustrated in Fig.~\ref{fig:linebands} for $z_{\rm obs}=0.25$,
where we plot $P(F)$ for various values of $\Gamma(z)$ by scaling the \cite{Haar12UVbackground} amplitude by
the factor $\alpha_\tau$, $\Gamma\to \Gamma/\alpha_\tau$. The value of $\alpha_\tau=0.7$ results in a very good fit to the observed curve for $F\le 0.7$. We note that there is a relatively large scatter in the measured values of $\tau_{\rm eff}$ for $z\le 1$ in the observations. We therefore suggest that it might be better to determine $\Gamma$ by fitting to $P(F)$, rather than by fitting to the single number $\tau_{\rm eff}$ at each given redshift.

We determined the parameters of the polynomial fit of  Eq.~(\ref{eq:poly_fit}) taking into account the observed evolution of $\tau_{\rm eff}$, the shape of the $P(F)$'s at different redshifts and, to a lesser extent, the amplitude of the flux power spectrum (discussed next). The relatively good agreement between the simulated and observed evolution of both $\tau_{\rm eff}$ and $P(F)$ for a chosen value of $\Gamma$ shows that it is possible to determine the required value of the ionising background that reproduces both types of observables - we do so in \S \ref{sect:UVB} below. We suggest that it is most likely that any remaining differences between simulated and observed $P(F)$'s at $F\ge 0.7$ for $z=0.25$ and $z=1$ are related to uncertainties in the continuum fitting of the data. However, the mismatch at $F\le 0.3$ at $z\approx 1$ remains unresolved. Notwithstanding that, we think that it is pleasing that the \eagle{} simulation is able to reproduce accurately the observed evolution of $\tau_{\rm eff}$ and $P(F)$.
We next show that this agreement extends to the flux power spectrum.

\subsubsection{The transmission power spectrum}
    \begin{figure}
     \includegraphics[width=\columnwidth]{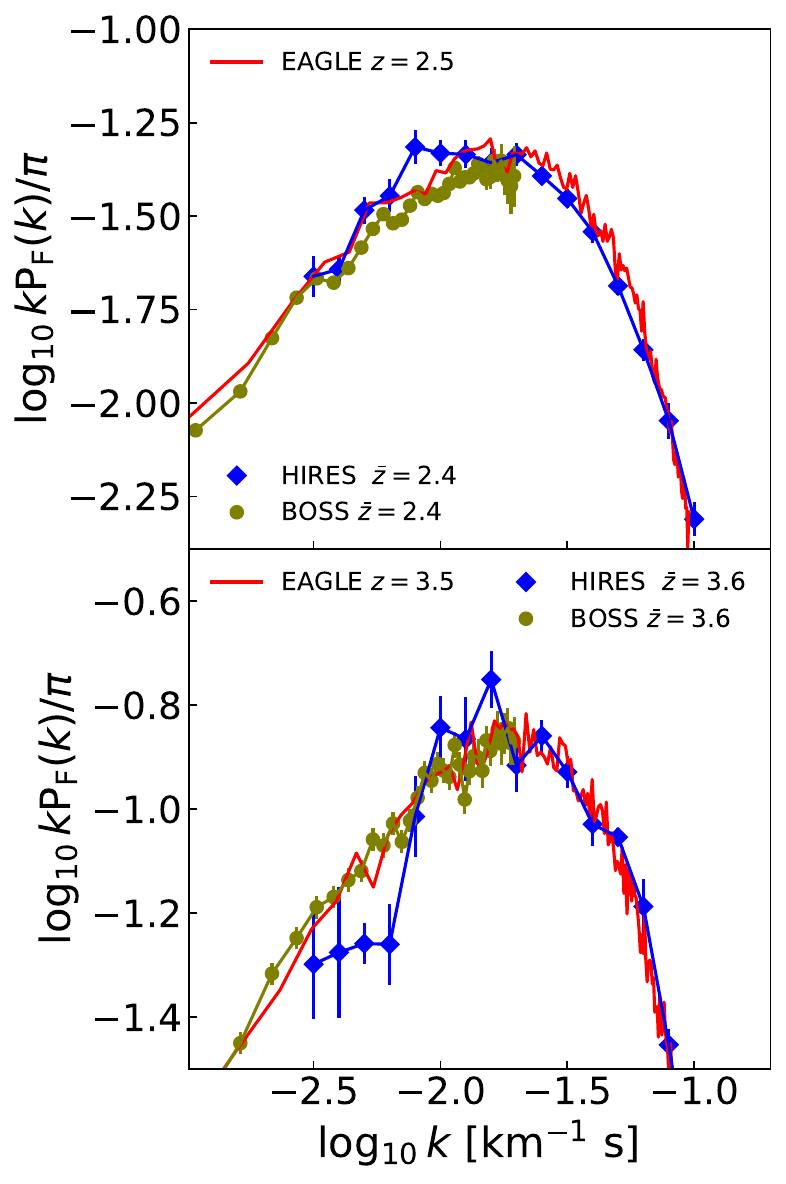}
      \caption{The \lya-forest flux power spectrum at redshifts of $z\approx 2.5$ ({\em upper panel}) and $3.6$ ({\em lower panel}). {\em Blue symbols with error bars} are from high-resolution observations with {\sc hires} on the {\sc keck} telescope (\protect\cite{Day19}), {\em golden symbols} are from the {\sc boss} lower-resolution spectra (\protect\cite{BOSS2013}). In both cases, $z_{\rm obs}$ is the mean observed redshift. The large fluctuations in the blue curves at $\log k\lesssim -1.5$ are due to sample variance,
      whereas the spectral resolution of the {\sc sdss} spectra limits their $k$-range to $\log k\lesssim -1.7$. The {\em red line} is from the \eagle\ L0100N1504 simulation, using the effective optical depth from Fig.~\ref{fig:fluxevol},
      $z_{\rm sim}$ is the redshift of the snapshot used. The \eagle\ spectra capture the rapid drop in the power spectrum for $\log k\gtrsim -1.8$ detected in the high-resolution data, as well as the power-law like shape of the power spectrum towards smaller $k$ detected in the {\sc sdss} data. The amplitude of the power spectrum is also reproduced.}
\label{fig:fsp_eagle}
\end{figure}

      

The transmission power spectrum, $P_F(k)$, is defined as
\begin{align}
    P_F(k) &= L\langle {\hat{\delta F}}(k)^\dag \hat{\delta F}(k)\rangle\,;\quad \delta F(v) \equiv \frac{F(v)}{\langle F(v)\rangle}-1\,,
\end{align}
where $\hat{\delta F}(k)$ is the Fourier transform of $\delta F(v)$, and $L$ is the length of the spectrum expressed in \vunit. $P_F(k)$ is commonly referred to as the \lq flux power spectrum\rq. The use of $P_F(k)$ to study the \lya-forest was pioneered by \cite{Croft98}. It has become a powerful tool which is used in a variety of studies, including constraining the nature of dark matter \cite[e.g.][and references therein]{Irsic24} and quantifying baryon-acoustic oscillations \cite[e.g.][]{Debulac15}.

We plot the dimensionless quantity $k\,P_F(k)/\pi$ in\footnote{
$k\,P_F(k)/\pi$ is the variance of $\delta F$ per $d\ln (k)$.} Fig.~\ref{fig:fsp_eagle} at $z\approx 2.5$ and $z\approx 3.6$. The blue data points are from high-resolution Echelle spectra, which can sample $P_F(k)$ to large values of $\log k\approx -1$ (here and below, we express $k$-values in units of \kunit). The relatively small number of available spectra results in significant sample variance for $\log k\leq -1.5$. The golden data points are from lower-resolution {\sc sdss} spectra that can sample $P_F(k)$ only for $\log k\lessapprox -1.7$: these data have much smaller sample variance for smaller $k$ compared to the Echelle spectra. Where the two data sets overlap, there is good agreement between them - given the large uncertainty due to sample variance in the blue data points.

The mock spectra from \eagle\ scaled to reproduce the effective optical depth from Fig.~\ref{fig:fluxevol} (red lines) fit the observations very well, from the largest $k$ to the smallest $k$. It would be interesting to examine to what extent feedback is required to get this level of agreement. The simulations in \cite{Bolton17} also reproduce the data 
from \cite{Viel04} (at $z=2.1$) and \cite{McDonald00} (at $z=3.9$), see their figure~7.

The shape of the power spectrum for $\log k\gtrapprox -1.5$ has been used to constrain the temperature of the absorbing gas \cite[e.g.][]{Theuns00, Molaro22}, and study the nature of the dark matter \cite[e.g.][]{Garzilli19, Irsic24}. The excellent agreement with the \eagle\ results show that there is only limited scope to improve the fit by changing the temperature of the gas or the nature of the dark matter.

\subsection{Spectra including many ions}
\begin{figure*}
     \includegraphics[width=\textwidth]{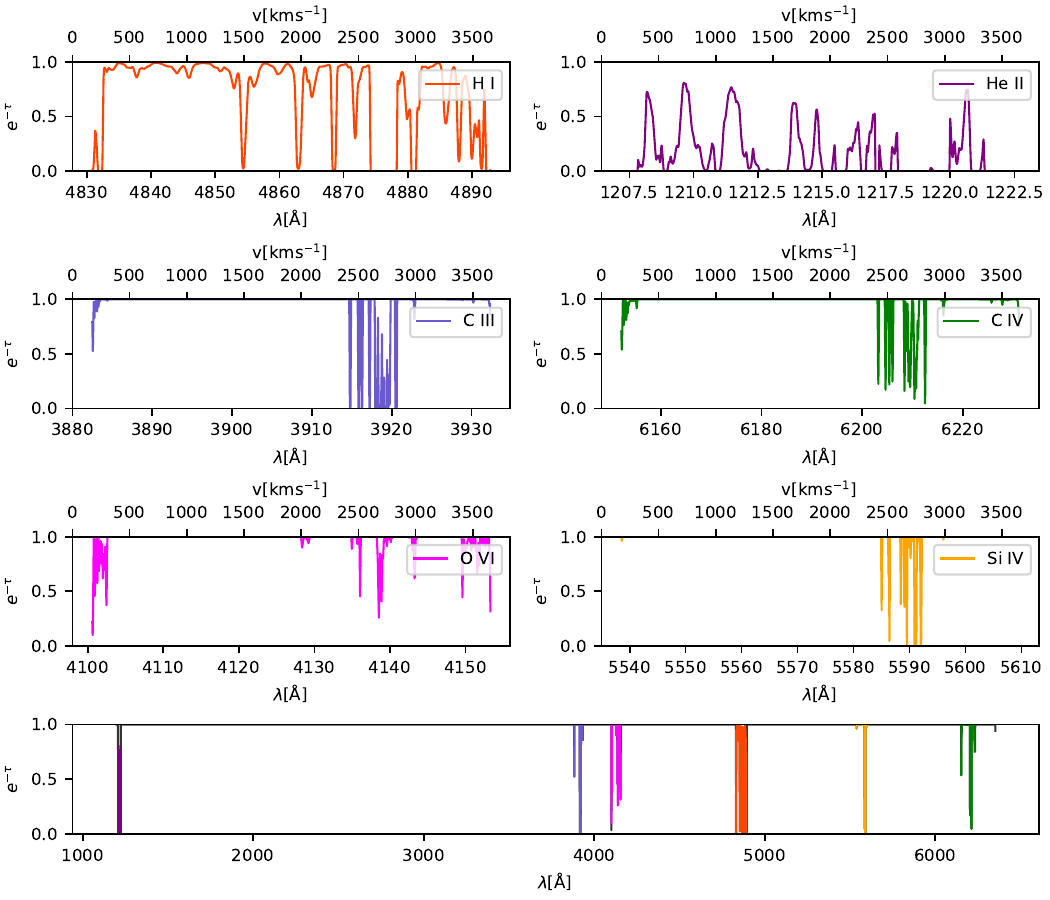}
      \caption{Example of the transmission, $\exp(-\tau)$, for several ions in a single sight line at $z\approx 3$ from the \eagle\ {\sc recal}L0050N1504 simulation. The simulation is post-processed with radiative transfer and the sight line picked so that it intersects a region of high density (here, at $v\approx 2700~\unit{km s^{-1}}$, see text). Different panels show different ions, as labelled in the legend. The lower horizontal axis is the observed wavelength, the upper horizontal axis is the Doppler velocity, $v$. The lower panel combines the stretches of spectrum in a single mock spectrum, with the colours repeating those of the ions in the upper panels.
      }
       \label{fig:shortspectra}
\end{figure*}

We illustrate the inclusion of other transitions in \spec\ using the higher resolution \eagle\ simulations, {\sc recal}L0050N1504. The snapshots of this simulation were post-processed with radiative transfer using the {\sc urchin} scheme described by \cite{Alta13revrt}. Briefly, {\sc urchin} propagates hydrogen ionising photons from low-density regions to higher-density regions and accounts for the self-shielding of hydrogen ionising photons in regions of high column density (i.e. it performs \lq inverse\rq\ radiative transfer). Self-shielding creates mostly neutral high-density regions embedded in a highly ionised lower-density Universe. Sight lines intersecting a high-density neutral region result in a \dla\ and/or \lls\ absorption and \cite{Altay11} showed that the statistics of these high column density systems obtained in the {\sc owls} suite \citep{Schaye10} agree well with observations. \cite{Rahmati15} found similarly good agreement with the data using the \eagle\ simulations.

Figure~\ref{fig:shortspectra} shows example spectra for a selection of ionic transitions for a single sight line through {\sc recal}L0050N1504 at $z=3$. The lower horizontal axis is the observed wavelength, $\lambda$, and the upper horizontal axis is the Doppler velocity, $v$ (which is the same for all spectra). This particular sight line was picked to intersect a high-density region, located at Hubble velocity $v\approx 2700~{\rm km~s}^{-1}$. Different panels show the strongest line of ions \ion{H}{i} 1215.67\AA, \ion{He}{ii} 303.78\AA, \ion{C}{iii} 977.02\AA, \ion{C}{iv} 1548.19\AA, \ion{O}{vi} 1031.91\AA, and \ion{Si}{iv} 1393.96\AA\ (where we state the laboratory wavelength, $\lambda_0$, of the line). We plot the strongest electronic transition for each ion, even for transitions that are doublets. For \ion{H}{i}, we plot $\exp(-\tau)$, where $\tau$ is the optical depth in the {\em absence} of a damping wing - this makes it easier to spot the high-density region in \ion{H}{i}.

There is strong \ion{H}{i} absorption at the location of the high-density region ($v=2700~{\rm km~s}^{-1}$), which coincides with strong absorption in \ion{C}{iii}, \ion{C}{iv}, \ion{O}{vi}, and \ion{Si}{iv}. There is little absorption from these ions at other velocities, except near $v=0$ where there is another, more modest, high-density region. As expected, there is significant absorption due to \ion{He}{ii} over the whole sight line - including at the location of the \dla.

The lower panel in Fig.~\ref{fig:shortspectra} illustrates how individual ionic transitions can be combined in a single mock quasar spectrum. We interpolate $\tau$ for each ion from the Doppler velocity of the individual sight line to the Doppler velocity of the mock spectrum. The colours repeat the colour of the ionic transition in the upper panels. In the next section, we show how we can generate a full mock quasar spectrum - including lines from all ions at all intervening redshifts.

\subsection{Mock absorption spectrum}
\begin{figure*}
     \includegraphics[width=\textwidth]{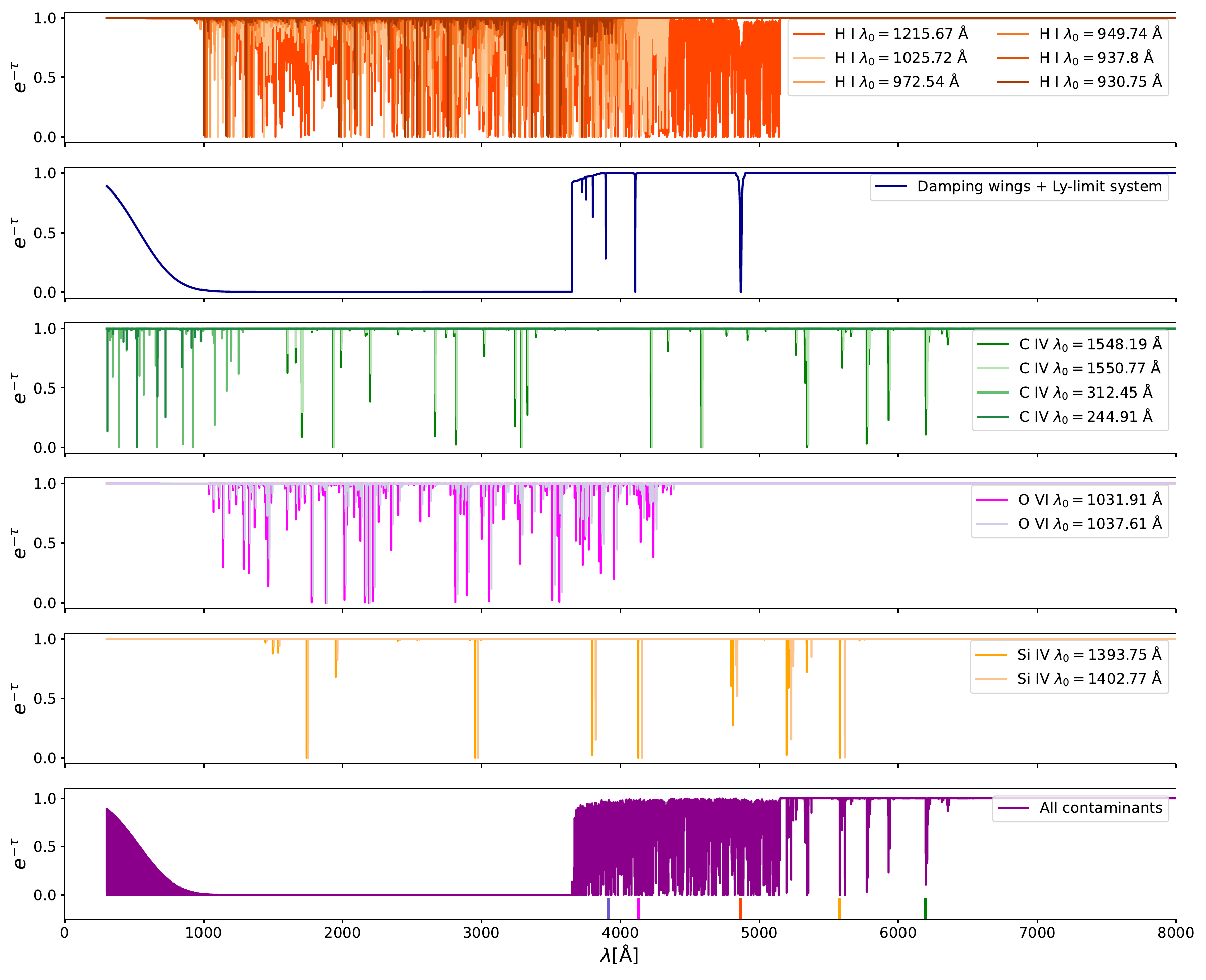}
      \caption{Example mock quasar spectrum for a redshift $z\approx 3.2$  quasar. Panels from top to bottom show some lines from the Lyman series of \ion{H}{i} (without damping wings), the \ion{H}{i} damping wings and bound-free (Lyman-limit) absorption, absorption due to \ion{C}{iv}, \ion{O}{vi} and \ion{Si}{iv}, and the absorption from all transitions combined. The spectrum contains the intervening \dla\ at $z\approx 3$ from Fig.~\ref{fig:shortspectra}; the short coloured lines show where this \dla\ imprints lines in the mock spectrum for the various ions in the upper panels using the corresponding colours. 
      }
      \label{fig:longspectra}
\end{figure*}

The spectra discussed in the previous section correspond to a single line of sight through an \eagle\ snapshot. These spectra are periodic since the simulation volume is periodic. Here we describe how we can combine many such sight lines to construct a full \lq observed\rq\ mock absorption spectrum, for which the \spec\ user specifies the wavelength range and the redshift of the background source, {\em e.g.} the quasar.

There are several ways in which we could accomplish this. One way is to make the sight line longer by shooting under some unconventional angle that allows the sight line to cross the periodic simulation volume several times without overlapping with itself. This can be accomplished relatively straightforwardly in the \eagle\ snapshot because it includes a hash table that allows easy identification of all particles in a given region of space. We note that there is no redshift evolution in such a spectrum since all particles are in the same snapshot. In addition, this method introduces artificial correlations in the spectrum.

Here we take advantage of another feature of \eagle: it contains \lq sight line\rq\ snapshots. These small snapshots are produced as the simulation is running: at frequent intervals, the running code selects 100 sight lines randomly through the simulation volume and outputs all particles that contribute to each of these 100 sight lines. We can now account for absorption by interlopers - and create a mock (quasar) spectrum - as follows. We start with a sight line file - or snapshot - close to background source's redshift, $z_{\rm qso}$, and compute a sight line spectrum. This sight line has some redshift extent, $\Delta z_1$. We now search for a sight line file that has redshift $z$ within some tolerance of $z_{\rm qso}-\Delta z_1$ and compute and append a sight line spectrum (with redshift extent $\Delta z_2$) for this sight line to the mock spectrum. The redshift $z$ of this sight line or snapshot is not necessarily identical to $z_{\rm qso}-\Delta z_1$: we scale the expansion factor by the small amount required before computing the spectrum. This small scaling affects the mean density, Hubble flow, peculiar velocity, and photo-ionisation rate. (We do not apply any scaling within a single sight line. Proper \lq light cone\rq\ output could be used if the simulation volume were much larger than that used here.) We then pick a sight line file with redshift within some tolerance of $z_{\rm qso}-\Delta z_1-\Delta z_2$. We repeat the procedure until we get to redshift $z=0$. For each individual sight line, we insert all lines due to all ions in the mock spectrum using the \id\ interpolation scheme.

An example output is shown in Fig.~\ref{fig:longspectra}. The top panel shows (some lines of) the Lyman-series of \ion{H}{i}, but without the damping wings and without continuum absorption. As expected, we see \lya{} absorption all the way from $\lambda=1215$\AA\ (corresponding to gas at $z=0$) to $\lambda=4860$\AA\ (corresponding to gas at $z=z_{\rm qso}$). The second panel from the top shows the damping wings and \ion{H}{i} continuum absorption only.  This particular sight line intersects the same $z\approx 3$ \dla\ that we analysed in the previous section - it is the origin of the damping wing and most of the continuum absorption. The lower panels show lines of \ion{C}{iv}, \ion{O}{vi} and \ion{Si}{iv} (including, this time, their doublet components). All these optical depths are combined to create the \lq full\rq\ mock spectrum in the lower panel. The small upward vertical lines show the locations where the \dla\ imprints ionic transitions in the mock spectrum. Absorption by the \dla\ is barely visible in the figure because the \lya{} forest itself is so dense, but the {\dla}'s corresponding continuum absorption causes almost complete absorption in the wavelength range $\lambda\in[1000, 3650]$\AA. The onset of continuum  absorption is spread-out in wavelength, this is due to the presence of bound-free absorption by high-column density absorbers with redshift slightly higher than that of the \dla\ (and to a lesser extent by thermal broadening). The transmission begins to recover below $\lambda\sim 1000$\AA.

We stress that this mock spectrum does not use light-cone data and that there are discontinuities in the spectrum where we stitched together individual spectra.

\subsection{Correlated absorption from \ion{H}{i} and \ion{Si}{iii}}
\begin{figure*}
     \includegraphics[width=\textwidth]{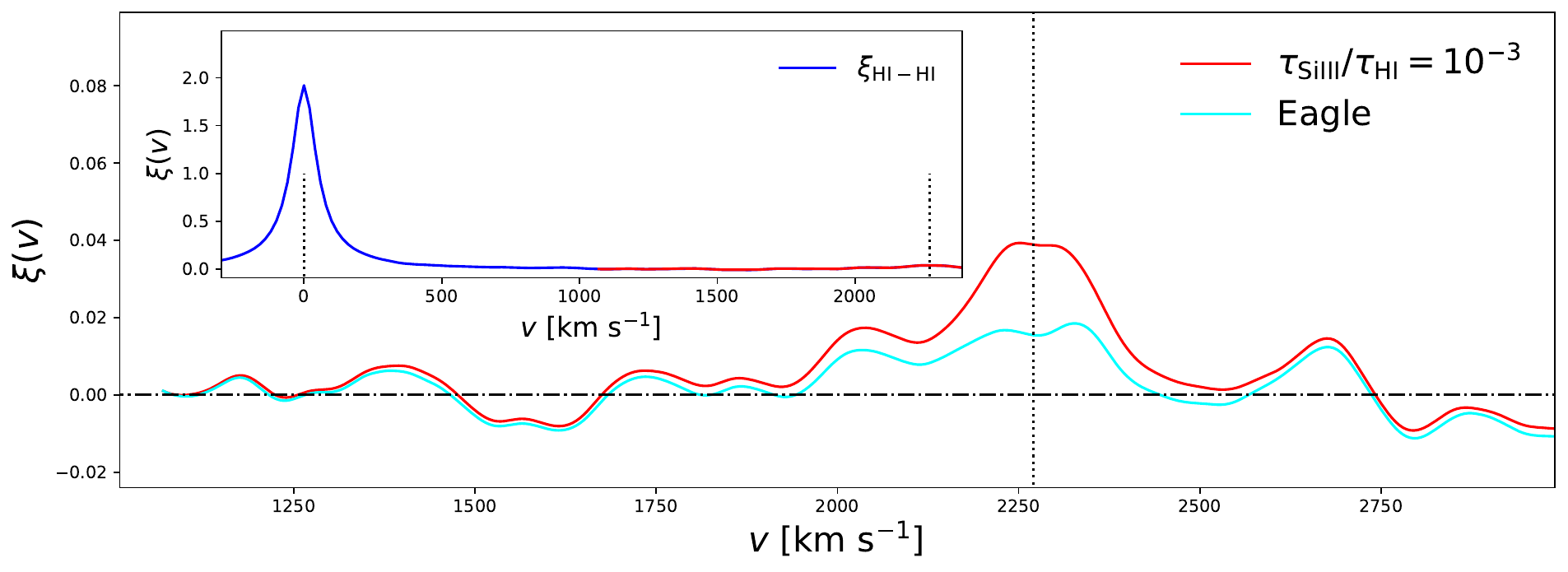}
     \caption{The feature in the absorption correlation function $\xi(v)$, Eq.~(\ref{eq:SiIII}), arising from correlated \ion{H}{i}-\ion{Si}{iii} absorption. The small inset shows $\xi(v)$ over a large velocity range, focussing on the peak at zero lag, $v=0$, indicated by a short vertical dotted line. The {\em blue line} is $\xi(v)$, the {\em red line} is the stretch of $\xi$ around $v_{\rm SiIII}$ where the \ion{H}{i}-\ion{Si}{iii} correlation generates a peak in $\xi$. The main panel focuses on $\xi$ around $v\sim v_{\rm SiIII}$, indicated by another short vertical dotted line. The {\em red line} is for a simulation where we arbitrarily set $\tau_{\rm SiIII}/\tau_{\rm HI}=10^{-3}$ everywhere. The {\em cyan line} is for the L0100N1504 \eagle\ snapshot at $z=3$ using the \protect\cite{Fauc09} ionising background. Wiggles in $\xi$ around $v\sim v_{\rm SiIII}$ are due to sample variance, but nevertheless there is a clear detection of correlated \ion{Si}{iii} absorption in \eagle.}
     \label{fig:SiIII}
\end{figure*}
As another example of the flexibility of \spec{}, we study {\em correlated} absorption between \ion{H}{i} and \ion{Si}{iii} in simulations. Such correlated absorption was identified in {\sc sdss} spectra by \cite{McDonald06}.

In general, the absorption from \ion{H}{i} and other ions at a given wavelength is mostly uncorrelated. For example, the \ion{H}{i} \lya\ absorption at $\lambda=4252~$\AA\ in Fig.~\ref{fig:longspectra} is imprinted by an absorber at $z=2.5$, but the \ion{O}{vi} absorption at the same observed wavelength is from $z=3.1$. Given the large distance between $z=2.5$ and $z=3.1$, any correlation between the physical absorbers that imprint these lines is small. When
we generate a mock spectrum as in the previous section, this correlation is actually zero, since we use {\em different} sight lines at these different redshifts. In general, this is not a serious limitation.

However, some ions have laboratory wavelengths, $\lambda_0$, which are very similar. In particular, \ion{H}{i} \lya\ has $\lambda_0=1215.67$~\AA, whereas \ion{Si}{iii} has a resonant transition at $\lambda_0=1206.5$~\AA. The small difference in laboratory wavelength should leave a detectable feature in the power spectrum. This feature is weak and requires averaging over many spectra to reduce sample variance. It was detected in the first release of {\sc sdss} spectra by \cite{McDonald06}, and the model they proposed to capture the correlation is now regularly employed when fitting \lya\ forest correlation functions or power spectra \cite[e.g.][]{duMas20, Palanque20}, see also \cite{Ma25}. For this pair of absorption lines, we use the {\em same} sight line to preserve the correlated absorption.

We quantify the strength of the correlation by computing the correlation function $\xi$, defined in the usual way as
\begin{align}
    \xi(v) &= \frac{\langle A(v')\,A(v'+v)\rangle}{\langle A(v)\rangle^2}-1\,;
    \quad A(v)\equiv 1-\exp(-\tau(v))\,,
    \label{eq:SiIII}
\end{align}
where $v$ is the Doppler velocity and the optical depth $\tau$ is due to \ion{H}{i} and \ion{Si}{iii} combined. Given the differences in laboratory wavelength, we expect $\xi$ to have a local maximum at velocity\footnote{The correlation function depends on the absolute value of $v$, but we compute it for both positive and negative values of $v$ in the simulations to quantify sample variance - hence the peaks at + and - 2270~km~s$^{-1}$.}
\begin{align}
    v_{\rm SiIII} &= \pm c \times 
    \left(
    2\frac{\lambda_{0, {\rm HI}}-\lambda_{0, {\rm SiIII}}}
    {\lambda_{0,{\rm HI}}+\lambda_{0,{\rm SiIII}}}-1
    \right)
    \nonumber\\
    &\approx \pm 2270~{\rm km~s}^{-1}\,.
    \label{eq:vSiII}
\end{align}

We plot the correlation function for $z\approx 3$ in Fig.~\ref{fig:SiIII}. The small inset focuses on the large peak in $\xi$ at zero lag ($v=0$, short vertical dotted line), resulting mostly from the \ion{H}{i}-\ion{H}{i} correlation at small velocity separation. The plot extends past $v=v_{\rm SiIII}$. The main panel focuses on the peak that results from the \ion{H}{i}-\ion{Si}{iii} correlation. We compare a toy model in which we simply set $\tau_{\rm SiIII}/\tau_{\rm HI}=10^{-3}$ everywhere (red line, also plotted in the inset) to the results obtained from \eagle\ using the \cite{Fauc09} ionising background. The toy model generates a peak in $\xi$ around $v=v_{\rm SiIII}$ (short vertical dotted line in both panels) that is detected even though the sample variance is large. There is also a clear detection of a peak in the correlation function computed from the simulation but at a somewhat lower amplitude than in the toy model.

\subsection{The surroundings of galaxies in absorption}
\begin{figure*}
     \includegraphics[width=\textwidth]{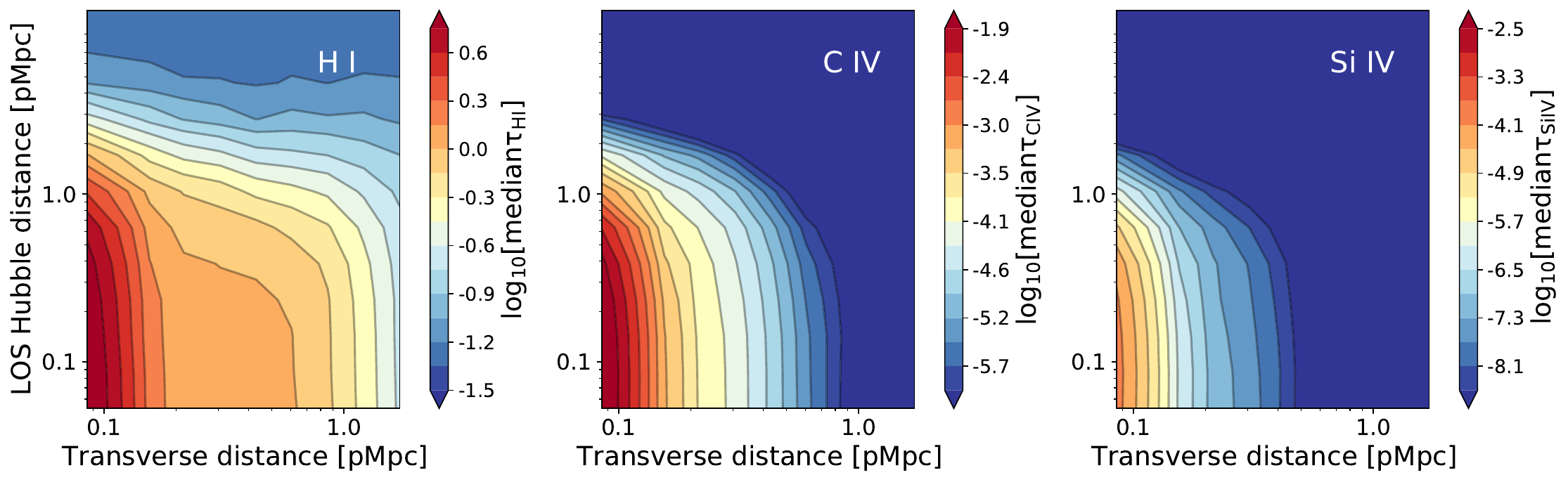}
      \caption{Ensemble-averaged two-dimensional absorption around haloes of mass $10^{12}M_\odot$ at $z\approx 3$ in the L0100N1504 \eagle\ simulation. The horizontal axis is the impact parameter of the sight line in proper units, the vertical axis is the Doppler velocity along the sight line, divided by the Hubble constant. The galaxies hosted by the halo has coordinates (0,0). Panels from left to right show the median \ion{H}{i} ($\lambda_0=1215.67$~\AA), \ion{C}{iv} ($\lambda_0=1548.19$~\AA) and \ion{Si}{iv} ($\lambda_0=1393.76$~\AA) optical depth. Absorption is stretched in the vertical direction due to the peculiar velocity of the absorbing gas.}
      \label{fig:turnerlikeplot}
\end{figure*}
There exists a large body of both observational and theoretical work on studying the surroundings of galaxies using absorption spectroscopy - both for local galaxies including the Milky Way, and for high redshift galaxies (see the Introduction for some references). 

As an example of how \spec{} can help interpret such data, we select haloes in the L0100N1504 \eagle\ simulation at $z=3$ with virial mass $\sim 10^{12}M_\odot$ from the \eagle {\sc sql} database \citep{McAlpine16}. We fire sight lines past each halo at a range of impact parameters and use \spec\ to compute a spectrum for various ions. For each impact parameter, we compute the median optical depth as a function of Doppler velocity $v$, setting $v=0$ at the location of the galaxy hosted by the halo. We then convert Doppler velocity to line-of-sight Hubble distance by dividing by the Hubble constant. 

Figure~\ref{fig:turnerlikeplot} shows the result for three selected ions; both axes are in proper units. Peculiar velocities distort the shape of the absorption from circular to extended in the redshift direction (the distortion is even more extreme once you realize that the vertical (redshift) axis extends to larger distances than the horizontal axis). 
Significant absorption in \ion{H}{i} extends to much larger distances (in both directions) than for the other elements, because C and Si are mostly ejected from the galaxy at the centre of the halo (with coordinates (0,0)) and its progenitors, whereas the hydrogen gas is mostly accreting from its surroundings. We also see that the \ion{C}{iv} distribution is more extended than the \ion{Si}{iv} distribution. It would be interesting to investigate whether that is due to the distribution of the elements, or caused by the different ionisation corrections. Plots such as Fig.~\ref{fig:turnerlikeplot} were analysed in much more detail by \cite{Turner17} (see their Fig.~3). These authors also account for observational effects (such as the finite spectral resolution leading to redshift uncertainties for the galaxies, and signal-to-noise limitations), and additionally compare to observations, finding good agreement between simulations and data.

\ifSubfilesClassLoaded{\bibliography{bibliography}{mybib}}
\end{document}

The evolution of the effective optical depth with redshift, $\tau_{\rm eff}(z)$, is plotted in Fig.~\ref{fig:fluxevol}. Observational data from 
\cite{Kim21} (for $z\le 3.5$) and \cite{Becker2013} (for $z\in [2,5]$) are shown, and fit by the polynomial
\begin{align}
   \tau_{\rm eff}(z) &=  A_0(1+z) + B_0(1+z)^2+ C_0(1+z)^3\,,
   \label{eq:poly_fit}
\end{align}
with parameters $ A_0 = 0.0433$, $B_0 = -0.0376$ and $C_0 = 0.0126$. 
The fit by \cite{Onorbe17} is shown as well.

We have scaled that amplitude of the \cite{Haar12UVbackground} at each snapshot of the {\sc eagle} simulation so that $\tau_{\rm eff}$ reproduces the polynomial fit of Eq.~(\ref{eq:poly_fit}). Most of the optical depth is due to lines with column density $N_{\rm HI}\in [10^{14}, 10^{15}]{\rm cm}^{-2}$. {\color{red} We should state what the evolution of the background is, compared to HM12.}

Using this amplitude of the ionising background, we have computed the histogram of transmissions in individial pixels for various redshifts.
These histograms are compared to the observed distributions from \cite{Kim21} in Fig.~\ref{fig:FPDF}.

 - show sample \lya-spectrum - and its decomposition in lines
 - discuss impact of lines on $\tau_{\rm eff}$
 - show flux pdf
 - show flux power spectrum

\subsection{Ions}
\begin{figure*}
     \includegraphics[width=\textwidth]{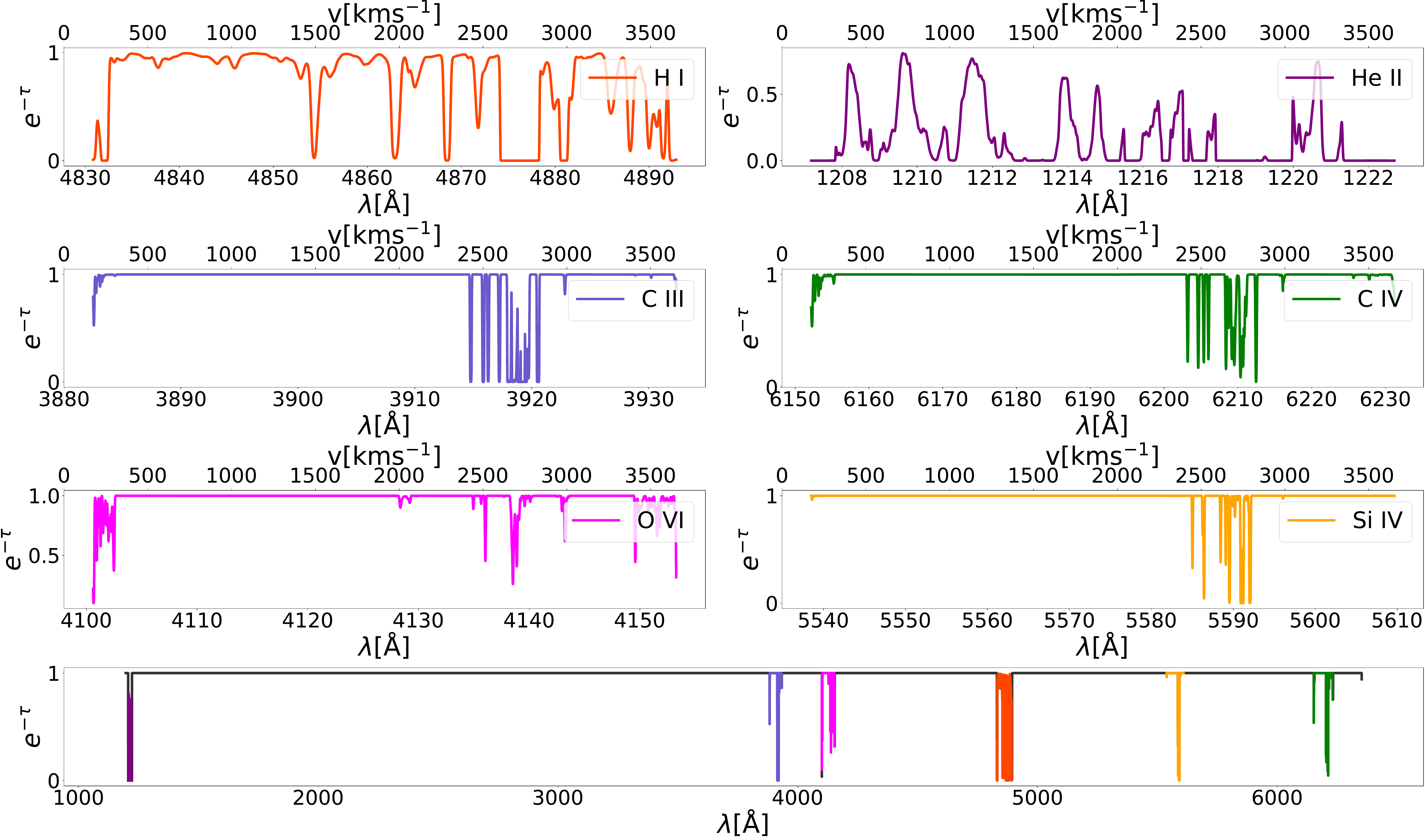}
      \caption{Example of the calculated transmission for a sightline at redshift z = 3? for different ions. In the horizontal axes we have both their position in the Hubble velocity and in their corresponding wavelength. }
      \label{fig:shortspectra}
\end{figure*}

\begin{enumerate}
    \item evolution of $\tau_{\rm eff}(z)$ for HM12 and for Ploeckinger, and our improved power-law fit

    \begin{equation*}
   \tau =  A_0(1+z) + B_0(+z)^2+ C_0(1+z)^3
    \end{equation*}
    
    with $ A_0 = 0.0433$,$B_0 = -0.0376$, $C_0 = 0.0126$
    {\color{red} TT to update with new suggestion for fitting function}
\begin{figure}
     \includegraphics[width=\columnwidth]{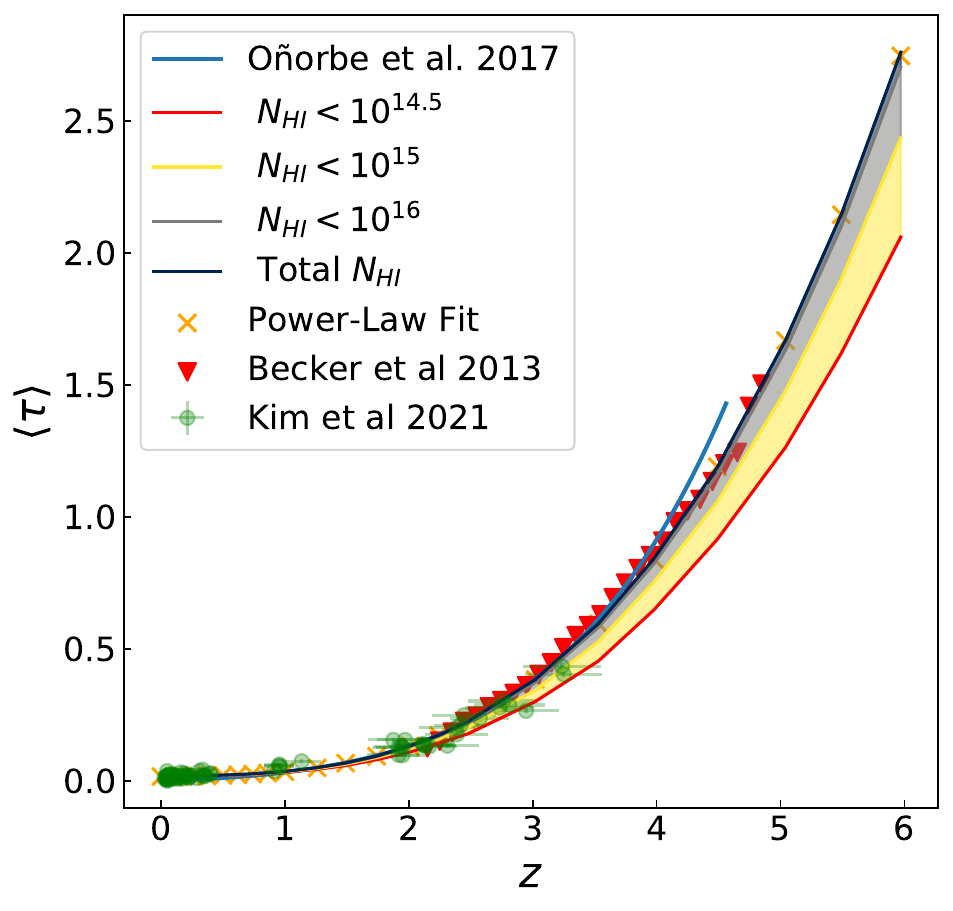}
      \caption{Where \protect\cite{Kim21}, \protect\cite{Becker2013}.}
\label{fig:fluxevol}
\end{figure}

    \item Gaussian decomposition for column density and equivalent width
    \item contribution to $\tau$ as a function of column density/equivalent width. Connect cddf to $\tau$
    {\color{red} Maybe move to other paper where we discuss other transitions as well}
    \item flux pdf at $z=2.5$ - pdf when transmission is replaced by square lines with correct equivalent width?
    {\color{red} Skip?}
    \item flux PS - convergence in Appendix (50 vs 100, $1500^3$ vs 756) - - pdf when transmission is replaced by square lines with correct equivalent width

    {\color{red} to do - plot nearly finished}
    \item CDDF (Gaussian decomposition or vpfit?)
    \item transmission around galaxies - compare to M Turner's plots (for more metals?)
    {\color{red} other section}
    
    \item impact of AGN feedback - also on the PS
\end{enumerate}

\noindent{Connecting $\tau_{\rm eff}$ to the \cddf.}\\

The column-density distribution function (hereafter \cddf) is defined as the number $N$ of lines with column density $N_{\rm HI}$ per unit absorption path, $dX$,
\begin{align}
    f(N_{\rm HI}) = \frac{{\rm d}^2N}{{\rm d}N_{\rm HI}{\rm d}X}\,.
\end{align}
The absorption path $dX$ is related to the proper path length, $dl$, and the Hubble velocity along $dl$, by
\begin{align}
    {\rm d}X=\frac{H_0(1+z)^3}{c}\,{\rm d}l = \frac{H_0(1+z)^3}{H(z)\,c}\,{\rm d}v\,,
\end{align}
where $H(z)$ is the Hubble parameter at redshift $z$. The absorption lines in $dX$ will absorb some fraction of the light, which will reduce the transmission to $\exp(-\tau_{\rm eff})$, where $\tau_{\rm eff}$ is called the effective optical depth. If we can assume that the absorption lines do not overlap with each other, then
the \cddf\ and the effective optical depth are realted by
\begin{align}
    1-\exp(-\tau_{\rm eff})
 &= \frac{H_0\,(1+z)^3}{H(z)\,c}\,\int_0^\infty f(N_{\rm HI})\,W(N_{\rm HI})\,{\rm d}N_{\rm HI}\,.
\end{align}
where
\begin{align}
    W(N_{\rm HI}) &= \int_{-\infty}^{\infty}\,\left(1-\exp(-\tau(N_{\rm HI})\right)\,dv\,,
\end{align}
is the equivalent width of the line in velocity units\footnote{$W$ is more commonly expressed in terms of the 
rest-frame equivalent width in \AA, $W_\lambda$. They are related by $W_\lambda=(\lambda/c)W$.}, and
$dX/dv=$
We note that $W$ depends on the line width, $b$, for saturated lines. For weak lines where $\tau\ll 1$, $W\approx \int_\infty^\infty \tau\,dv=\sigma_\alpha\,N_{\rm HI}$, does not depend on $b$, and

Relate power spectrum to power-law profile of the absorbers, as in \cite{Ma00}\\

SIV absorbers: see \cite{dodorico22} and \cite{Cooksey11}
For analysis using Eagle, see \cite{Rahmati16}

\subsection{SiIII}
McDonald (06) defines the impact of SiIII by the parameter $a$, where the over density $F(v)/\bar F-1\equiv \delta_F(v)$ in the spectrum with SiIII is related to the Hydrogen-only over density $\delta(v)$, by $\delta_F(v)=\delta(v)+a\delta(v+v_3)$, where $v_3/c=\lambda_{\rm SiIII}/\lambda_{HI}-1$ is the velocity offset between the two lines. The correlation functions are then related by
\begin{align}
\xi_F(\Delta) &\equiv \langle \delta_F(v)\,\delta_F(v+\Delta)\rangle\nonumber\\
              &=(1+a^2)\xi(\Delta)+a\xi(\Delta+v_3)+a\xi(\Delta-v_3)\,,
\end{align}
(McDonald06, their Eq.~(14).

A simple physical to capture the correlation is to assume that optical depths of the two ions are proportional, $\tau_{\rm SiIII}=\alpha\tau_{\rm HI}$, for some constant $\alpha$. The total optical depth at velocity $v$ is then $\tau_{\rm tot}(v)=\tau_{\rm HI}(v)+\alpha\tau_{\rm HI}(v+v3)$. We can then define two correlation functions,
\begin{align}
    1+\xi_{\rm HI}(\Delta) &\equiv \bar F_{\rm HI}^{-2}\,\langle \exp(-\tau_{\rm HI}(v))\,\exp(-\tau_{\rm HI}(v+\Delta))\rangle\nonumber\\
    1+\xi_{\rm tot}(\Delta) &\equiv \bar F_{\rm tot}^{-2}\,\langle \exp(-\tau_{\rm tot}(v)\,\exp(-\tau_{\rm tot}(v+\Delta))\rangle\nonumber\,.
\end{align}
Here, $\xi_{\rm HI}$ is the correlation function of the transmission for $\alpha=0$ (no Si~III),
$\xi_{\rm tot}$ is the correlation function in the presence of Si~III, and $F_{\rm HI}$ and $F_{\rm tot}$ are the mean transmission in these two cases, $F_{\rm HI}\equiv \langle\exp(-\tau_{\rm HI}(v)\,\rangle$ and $F_{\rm tot}\equiv \langle \exp(-\tau_{\rm tot}(v)\rangle$. The HI-SiIII correlation creates a bump in $\xi_{\rm tot}(\Delta)$ at $\Delta=\pm v_3$, the amplitude of which follows from
\begin{align}
    \bar F^2_{\rm tot}&\big(\xi_{\rm tot}(\Delta=\pm v_3)+1\big)
    \nonumber\\
    &=\big\langle
    \exp\left(-\tau_{\rm HI}(v)-\alpha\tau_{\rm HI}(v+v_3)-\tau_{\rm HI}(v+v_3)-\alpha\tau_{\rm HI}(v+2v_3)\right)
    \big\rangle\nonumber\\
    &\approx
    \big\langle\exp\left(-\tau_{\rm HI}(v)\right)\big\rangle\,
    \big\langle\exp\left(-\tau_{\rm HI}(v+v_3)-\alpha\tau_{\rm HI}(v+v_3)\right)\big\rangle\nonumber\\
    &\times \big\langle\exp\left(-\alpha\tau_{\rm HI}(v+2v_3)\right)\big\rangle\nonumber\\
    &\approx \bar F_{\rm HI}\times\bar F_{\rm tot}\times 1\,.
\end{align}
We made the following approximations: ({\em i}) at a separation $\Delta=v_3$, the correlation between the HI optical depths can be neglected,
({\em ii}) the second ensemble average equals $\bar F_{\rm tot}$, and ({\em iii}) $\alpha\ll 1$ so that the last ensemble average equals 1. The final results is that
\begin{align}
    \xi_{\rm tot}(\Delta=v_3) \approx \frac{\bar F_{\rm HI}}{\bar F_{\rm tot}}-1\,.
\end{align}

A simple physical model to describe this correlation 

If we write instead $F_t(v)=F(v)+b\,F_t(v+v_3)$ (which makes much more sense), for which
$\bar F_t=(1+b)\bar F$, and
\begin{align}
\langle F_t(v)F_t(v+\Delta)\rangle &= (1+b^2)\langle F(v)F(v+\Delta)\rangle\nonumber\\
&+ b\langle F(v)F(v+v_3+\Delta)\rangle\nonumber\\
&+ b \langle F(v)F(v-v_3+\Delta)\rangle\,.
\end{align}
The correlation functions are then related by
\begin{align}
(1+b)^2\,(1+\xi_t(\Delta)) &= (1+b^2)(1+\xi(\Delta))\nonumber\\
&+b(1+\xi(\Delta+v_3))\nonumber\\
&+b(1+\xi(\Delta - v_3))\,.
\end{align}

Let $F_{\rm HI}$ and $F_{\rm SiIII}$ be the mean transmisison due to HI and SiIII seperately.
The ratio of optical depths is then approximately
\begin{align}
    f &=\frac{1-F_{\rm SiIII}}{1-F_{\rm HI}}\approx \frac{\tau_{\rm SiIII}}{\tau_{\rm HI}}\,,
\end{align}
so that
\begin{align}
F_{\rm SiIII} &= (1-f) + f F_{\rm HI}\,,
\end{align}

and the spectrum is $F(v) = F_{\rm HI}(v)+aF_{\rm HI}(v-v_3)$

SiIII will then generate a bump in the correlation function at velocity separation $\Delta$
\begin{align}
(\bar F)^2(1+\xi(\Delta)) &= \langle F(v) F(v+\Delta)\rangle\nonumber\\
&=\langle (F_{\rm HI}(v)+aF_{\rm HI}(v-v_3))\,(F_{\rm HI}(v+\Delta)+aF_{\rm HI}(v+\Delta-v_3))\rangle\nonumber\\
&= (\bar F_{\rm HI})^2\,\left((1+a^2)(1+\xi(\Delta)) + a (1+\xi(\Delta-v_3))+a(1+\xi(\Delta+v_3)\right)\,.
\end{align}

\subsection{Transmission around galaxies}

\subsection{Longspectra}

    SpecWizard posses the tool $\verb|SpecWizard.LongSpectra|$ that simulates a spectrum getting from a spectrograph with a range of redshifts. For this we define the wavelength range $\{\lambda_{min},\lambda_{max}\}$ with some pixel size $\Delta \lambda$ and the redshift in which the quasar is located $z_{q}$. The components of the spectra is extracted from simulation output that are within the redshift range of the spectra. Our wavelength range has an equivalent velocity range given by:
    \begin{equation}
        \frac{d\lambda}{\lambda} = \frac{dv}{c}
    \end{equation}
    A simulation volume will cover some redshift range $\{z_{sim},z_{sim}+\Delta z\} $ where $\Delta z$ is given by extend of such volume. Depending on the specific line we want to include into our longspectra there will be a rest wavelength $\lambda_{l,r}$. Thus the location of the line in wavelength will be $\{\lambda_{l,i},\lambda_{l,f}\} = \{\lambda_{l,0}(1+z_{sim}),\lambda_{l,0}(1+z_{sim}+\Delta z)\}$ and velocity space
    \begin{equation}
        v = c \ln{\frac{\lambda_{l}}{\lambda_{min}}}
    \end{equation}

\begin{figure*}
     \includegraphics[width=\textwidth]{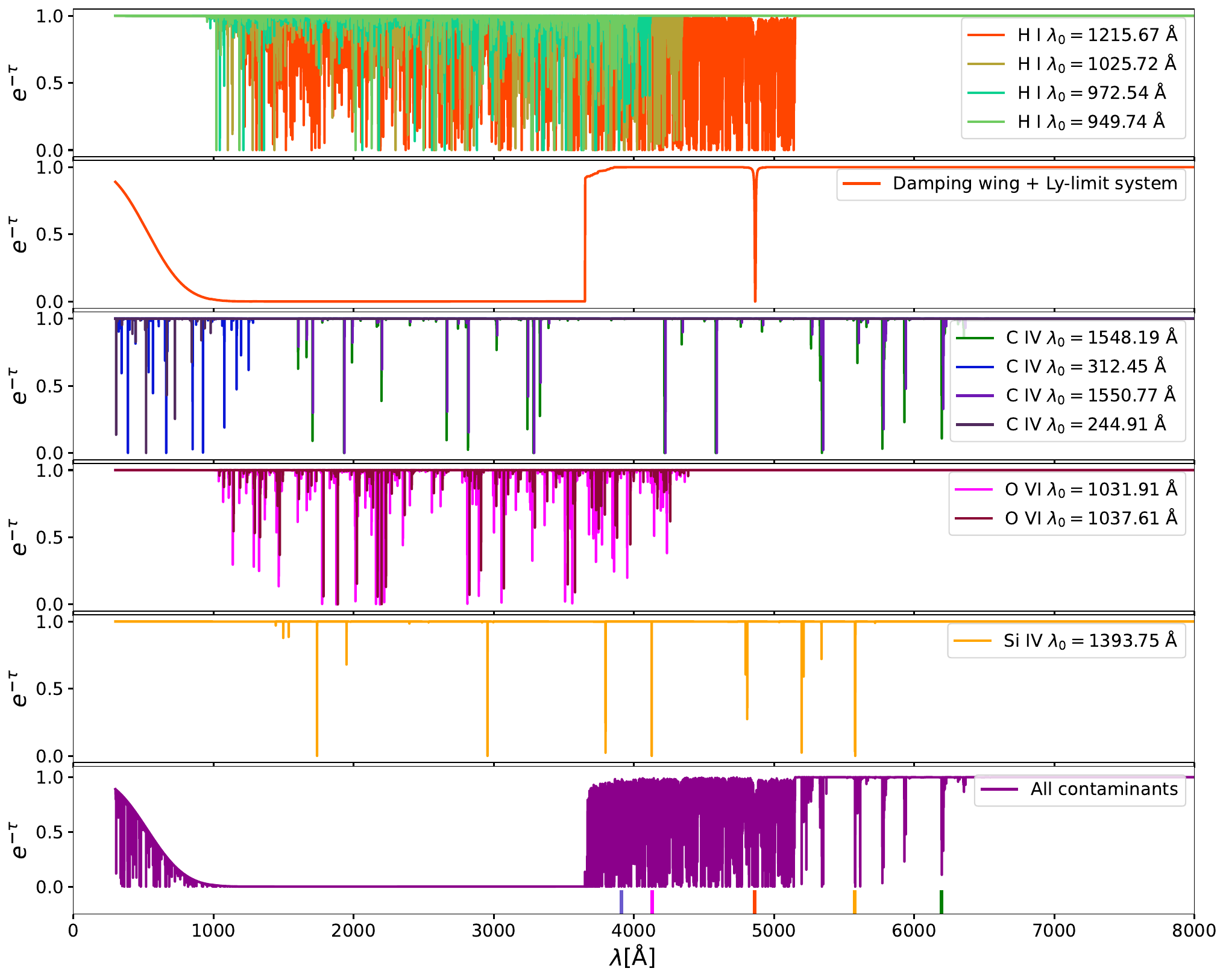}
      \caption{Example of the calculation of the longspectra with the ly-alpha series + Damping wings and LL-systems. We include ions that contribute to the spectrograph and some of other transitions. In the bottom we include all the contamination. In lines we indicate the shortspectra from Figure X that was introduced to be part of the longspectra }
      \label{fig:shortspectra}
\end{figure*}

\subsection{Metal absorbers}

\begin{enumerate}
    \item evolution of $\tau_{\rm eff}(z)$ for both. Connection to evolution of $\tau_{\rm eff}(z)$ for HeII and connection to HeII reionisation
    \item CIV and SiIV CDDF
    \item rise and fall of CIV
    \item, CIV-HI cross correlation
    \item impact of the feedback on the metal cddf
    \item correlation with galaxies
     \item quasar spectrum and contamination
\end{enumerate}

\ifSubfilesClassLoaded{%
  \bibliography{bibliography}%
}{mybib}

\end{document}

\section{The evolution of the photo-ionisation rate}
\label{sect:UVB}
\begin{figure}
     \includegraphics[width=\columnwidth]{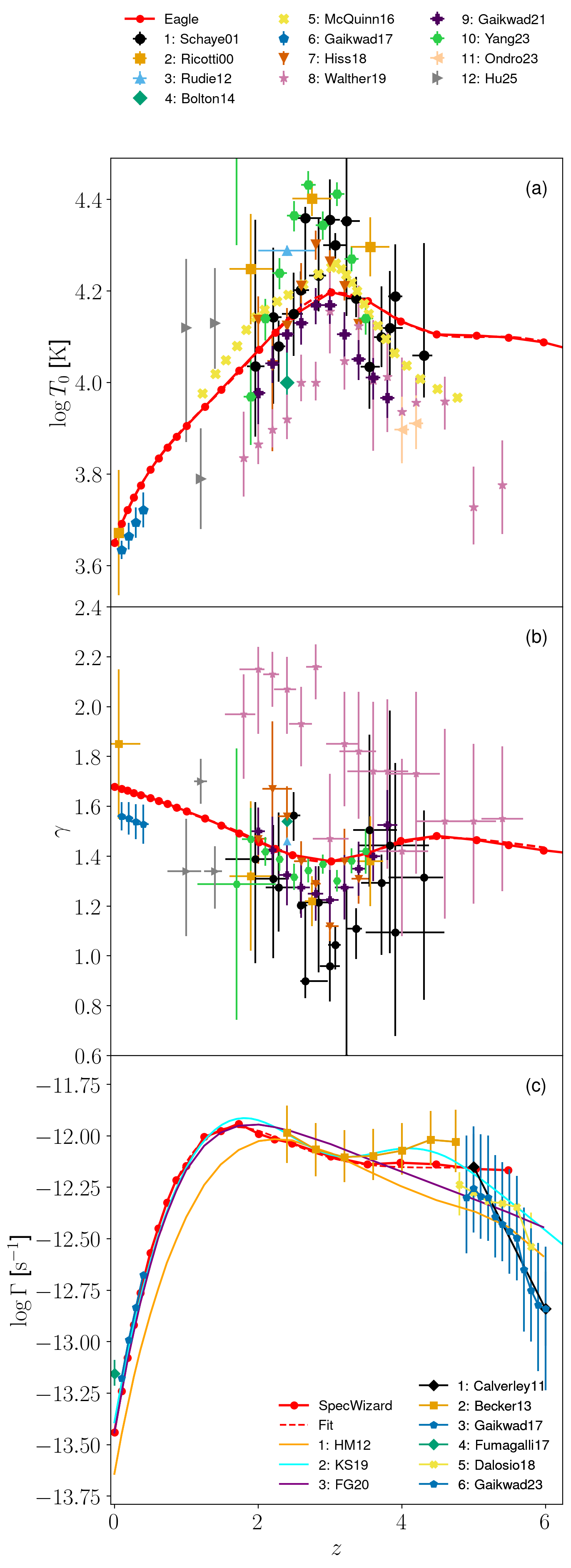}
      \caption{Upper two panels: evolution of the parameters $T_0$ (panel a) and $\gamma$ (panel b) of the temperature-density relation, $T=T_0~(\rho/\bar\rho)^{\gamma-1}$. The simulated evolution is shown as the red line.
      Lower panel: panel (c) shows the evolution of the photo-ionisation rate, $\Gamma$, inferred by comparing mock \spec{} spectra to \lya{}-forest data, as a red line.  In all panels, coloured curves and symbols show measurements from the literature, as per the caption above the figure, for panels (a) and (b), and inside the figure, for panel (c). See text for details and discussion. The {\em dashed red line}, mostly hidden by the red line, are fits to $T_0$, $\gamma$ and $\Gamma$,
      from Eqs.~(\ref{eq:Tfit}-\ref{eq:Gfit}).} 
\label{fig:Temperature}
\end{figure}

We demonstrated in section \ref{sect:Lya} that mock \lya{}-forest spectra from the \eagle{} simulation reproduce the observed evolution of the effective optical depth, flux {\sc pdf} and flux power spectrum, provided we allow ourselves the freedom to scale the imposed value of the photoionisation rate, $\Gamma$, from the default values taken from \cite{Haar12UVbackground}. These statistics also depend on the temperature of the absorbing medium. In this section, we show that the \eagle{} simulation reproduces the observed evolution of the temperature as well, allowing us to measure the evolution of $\Gamma$.

Around the mean density, $\bar\rho$, of the \igm{}, the photo-heated gas that dominates \lya{} absorption in simulations follows a well-defined relation between temperature, $T$, and density, $\rho$, of the form\footnote{The exponent is traditionally denoted by $\gamma$, but is generally not equal to the adiabatic exponent 5/3 of the gas.} $T(\rho)=T_0\,(\rho/\bar\rho)^{\gamma-1}$, where $T_0$ is the temperature at the mean density \citep{Hui97EOSIGM}, see also reviews by \cite{Meiksin09, McQuinn16}.  Gas with temperature below $T(\rho)$ is quickly photo-heated onto this temperature-density relation. However, gas with temperature above the $T(\rho)$ relation, maybe as a result of shocks from structure formation or feedback from star formation or {\sc agn}, does not cool back onto the relation since the cooling time is generally much longer than the Hubble time. Nevertheless, the majority of the gas around the mean density follows the relation  $T(\rho)=T_0\,(\rho/\bar\rho)^{\gamma-1}$ tightly. $T_0$ and $\gamma$ evolve with $z$ due to the effects of photo-heating and adiabatic expansion \cite[e.g.][]{Sanderbeck16}.

We have measured the evolution of $T_0$ and $\gamma$ in the \eagle{} simulations as follows. We bin the simulation
particles in bins of gas density for a snapshot at a given redshift in logarithmic density bins of 0.02~dex, determine the 20 per cent lowest temperature in each density bin, and then fit for density and 20$^{\rm th}$ percentile temperature
with the power law parameters $T_0$ and $\gamma$ for density bins within 0.2~dex from $\bar\rho$. Repeating this process for all snapshots yields $T_0(z)$, and $\gamma(z)$, shown in panels (a) and (b) of Fig.~\ref{fig:Temperature} by a solid red line; solid circles denote the redshifts of individual \eagle{} snapshots. The $T-\rho$ relation in \eagle{} is well defined and well-fit by the power-law, therefore the inferred values of $T_0$ and $\gamma$ are not sensitive to the exact choice of density bins or temperature percentile.

Panels (a) and (b) also show a selection of measured values collected from the literature as coloured symbols. As per the caption above the figure, these are: 1: \cite{Schaye00}, 2: \cite{Ricotti00}, 3: \cite{Rudie12}, 4: \cite{Bolton14}, 5: \cite{McQuinn16} (themselves taken from \cite{Upton15} and based on \cite{Becker11} and \cite{Boera14}), 6: \cite{Gaikwad17}, 7: \cite{Hiss18}, 8: \cite{Walther19}, 9: \cite{Gaikwad21}, 10: \cite{Yang23}, 11: \cite{Ondro23} and 12: \cite{Hu25}. We note that not all data have associated error bars, and not all authors determine $\gamma$. In most cases, $T_0$ and $\gamma$ are determined by measuring one or more statistics of \lya{} absorption in quasar spectra, and comparing these to simulations analysed in the same way. The original method of \cite{Schaye99} applied in \cite{Schaye00} involves fitting the \lya-forest lines with Voigt profiles. Plotting the logarithm of the line-width $b$ versus that of the column density, $N_{\rm HI}$, for each line reveals a cut-off in $b$ at given $N_{\rm HI}$, which is affected by the temperature of the absorbing gas. Having measured that cut-off, simulations are used to relate the parameters of the power-law cut-off to $T_0$ and $\gamma$. The analysis of 1, 3, 4, \& 7 measure the $b-N_{\rm HI}$ cut-off, 10 fits the $b-N_{\rm HI}$ cut-off using the method described by \cite{Garzilli15, Garzilli20}, 2 \& 12 fit the $b-N_{\rm HI}$ distribution, 5, 6 \& 11 use the curvature of absorption lines, 
analyses 6 \& 8 use the flux pdf and/or the flux power spectrum, whereas analysis 9 uses all of these, including the wavelet method described by \cite{Zaroubi00}.

In the data, $T_0$ increases with redshift, reaching a peak around $z\sim 3$, and then decreases towards higher $z$. The sharpness of the peak and the value of $T_0$ at the peak, differ significantly between measurements, with \cite{Yang23} preferring a narrow high peak, and the data by \cite{Walther19} falling below the general cloud of data points by 0.1-0.2~dex. The latter dataset also yields significantly higher values of $\gamma$. The \eagle{} simulation also shows a peak in $T_0$ and associated trough in the value of $\gamma$, which are both due to extra energy injected by the reionisation of \ion{He}{ii} to \ion{He}{iii}. Such an injection of energy is required to explain the unexpected large line widths in the data, as originally suggested by \cite{Theu98P3MSPH}.

We note that the different methods shown in Fig.~\ref{fig:Temperature} may not yield the same value of $T_0$ when there are spatial fluctuations in $T_0$. Such fluctuations are expected to be present at $z\sim 3-4$ when \ion{He}{ii} is being reionised by relatively rare sources. Methods based on a cut-off are using the {\em narrowest} lines and hence probe the lowest values of $T_0$. Methods using
the flux {\sc pdf} or power spectrum may yield a weighted value of $T_0$, 
resulting in a higher value of $T_0$ than obtained from the $b-N_{\rm HI}$ cut-off. 
Analysis 8 uses the flux power spectrum - but, if anything, the inferred values of $T_0$ are {\em lower} than inferred from the cut-off. \cite{Theuns02} searched explicitly for temperature fluctuations in the data, finding no evidence for variations in $T_0$ larger than 50 per cent over scales of $5000~\unit{km s^{-1}}$. The impact or indeed presence of temperature fluctuations is currently not well constrained.

The \eagle{} simulation traverses the general cloud of data points below $z=4$, and is also consistent with the measurements at $z\sim 0$. Above $z\sim 4$, $T_0$ remains approximately constant in \eagle{}. In contrast, there is a hint in the measurements of 5 \& 11 that $T_0$ falls rather steeply from $z=3$ towards $z=6$. Such a fall is also suggested by dataset 8, however its values of $T_0$ are, in fact, low at all $z$.

In summary: there is significant scatter, of order 0.2~dex, as well as bias between currently measured values of $T_0$ and $\gamma$. Given the relatively large uncertainty in the measurements, the temperature evolution in \eagle{} is consistent with the measurements, at least in the redshift range $z\in [0, 4]$.

In \eagle{}, the temperature approaches a slowly varying asymptotic value for $z\to 0$, which can be fit well by an exponential. Superimposed on this global trend is a localized increase in $T_0$, well described by a Gaussian. The exponent $\gamma$ exhibits a predominantly monotonic evolution with $z$, with a localized dip and asymmetric curvature at a similar redshift to where $T_0$ increases. We fit these shapes to the evolution of $T_0$ and $\gamma$ in \eagle{}, 
\begin{align}
\label{eq:Tfit}
    \log T_0(z) &= \log T_\infty - a\,e^{-z/\tau} + A \exp\left(-\frac{(z - z_0)^2}{2\sigma^2}\right)\\
    \log T_\infty &= 4.04060,\quad
a = 0.38724,\quad
\tau = 1.08114\nonumber\\
A &= 0.16302,\quad
z_0 = 3.23385,\quad
\sigma = 1.14107\nonumber\\
\label{eq:gfit}
\gamma(z) &= 
\gamma_\infty + \frac{\gamma_0 - \gamma_\infty}{1 + (z/z_c)^\alpha}
+ A_1 \exp\left(-\frac{(z - z_1)^2}{2\sigma_1^2}\right)\nonumber\\
&+ A_2 \exp\left(-\frac{(z - z_2)^2}{2\sigma_2^2}\right)\\
\gamma_0 &= 1.52885,\quad
\gamma_\infty = 1.21574,\quad
z_c = 1.23847\nonumber\\
\alpha &= 1.65165\nonumber\\
A_1 &= -0.13526,\quad
z_1 = 2.87167,\quad
\sigma_1 = 0.74298\nonumber\\
A_2 &= 0.24204,\quad
z_2 = 3.67962,\quad
\sigma_2 = 3.59227\,.
\end{align}
These fits are shown as red dashed lines in panels (a) and (b): they mostly overlap with the red line.

Given the agreement with the measured evolution of $T_0$ and $\gamma$, we can use the simulation to infer the amplitude of the ionising background, $\Gamma$, which is plotted as the red line in panel (c) of Fig.~\ref{fig:Temperature}, with solid circles corresponding to the redshifts of individual snapshots. We use the measured value of the scaling factor $\alpha$ from Fig.~\ref{fig:tau_scaling}, with the scaling starting from the \cite{Haar12UVbackground} \uvb{}. The dashed red line, which mostly falls on top of the red line, is a fit which uses a Gaussian to fit the peak (parameters $A$, $\mu$ and $\sigma$), an exponential to fit the rapid rise from $z=0\to 1$ (parameters $B$ and $k$), and a constant ($C$) to fit the asymptotic behaviour at $z>4$,
\begin{align}
    \log\Gamma(z)[\unit{s}^{-1}] &= A\,\exp(-\frac{(z-\mu)^2}{2\sigma^2}) + B\exp(-kz) + C\nonumber\\
    A&=0.4664, \mu=1.005,\sigma=1.064\nonumber\\
    B &=-1.592,k=1.301\,;\quad  C =-12.15\,,
    \label{eq:Gfit}
\end{align}
valid for $z\in [0, 4.5]$. 
According to the model of \cite{Schaye01}, the column density of a line is proportional
to $T_0^{-0.26}\,\Gamma^{-1}$ (their equation 10). Given the relatively large uncertainty in the value of $T_0$, we suggest that the error on
$\Gamma$ is dominated by that in $T_0$. Using the error on $\log T_0$ of 0.1-0.2~dex from earlier, we estimate the uncertainty in $\log\Gamma$ to be 0.26 times that on $\log T_0$, therefore 0.03-0.05~dex.

Panel (c) also shows published models of the \uvb{} as lines, and measured values as symbols with error bars, as per the legend. The models are 1: \cite{Haar12UVbackground}, 2: \cite{Khaire19} \& 3: \cite{Faucher20}. The data are 1: \cite{Calv11UVbg}, 2: \cite{Becker2013}, 3: \cite{Gaikwad17}, 4: \cite{Fumagalli17}, 5: \cite{DAlo18UVbg} \& 6: \cite{Gaikwad23}. As with the determination of $T_0$ and $\gamma$, measurements based on the data are compared to mock spectra to infer $\Gamma$, except for the data point of \cite{Fumagalli17} which is determined from H$\alpha$ fluorescence of the outskirts of a nearby, edge-on, galaxy disk. We note that the model of \cite{Khaire19} is designed to fit the increase in $\Gamma$ between $z=3$ and $z=4$ that is suggested by the data of \cite{Becker2013}. 

We find excellent agreement between our measured value of $\Gamma$, the model of \cite{Faucher20}, and the measurements of \cite{Fumagalli17} and \cite{Gaikwad17} from $z\sim 0$ to $z=2$, namely $\Gamma$ increases by more than an order of magnitude from $z=0$ to $z=2$. Such a rapid increase was shown to be necessary to reconcile the $z=2$ \lya-forest with the $z\sim 0$ \lya-\lq savannah\rq\ by \cite{Theuns98a}. From $z=2$ to $z=4$, we find very little evolution in the value of $\Gamma$, as do the measurements of \cite{Becker2013}, and all three models (with \cite{Haar12UVbackground} predicting a value of $\Gamma$ lower by $\approx 0.1$~dex). Models, data, and our inferred value then seem to diverge from $z=4$ to $z=5$, with the values of \cite{Becker2013} increasing, the models decreasing, and our value remaining almost constant. Above $z\sim 5$, datasets 1 and 6 decrease rapidly with increasing $z$, 5 remains approximately constant, and the models decrease slowly. If $T_0$ indeed decreases rapidly above $z=4$, as suggested by the data in panel (a), then we would need a {\em larger} value of $\Gamma$ in the simulation to be consistent with the \lya{} flux statistics. 

We summarise as follows: the \eagle{} simulation follows the general cloud of points for the measured evolution of $T_0$ and $\gamma$, at least over the redshift range $z\in [0, 4]$. Some published data are at face value inconsistent with the general trend. For this evolution of $T_0$ and $\gamma$, the inferred evolution of the photoionisation rate $\Gamma$ is in excellent agreement with other measured values. The agreement is of order 0.1~dex for $z\in [4, 5]$, and better at lower $z$. Above $z=5$, different data sets do not agree well with each other. We therefore suggest that the value of $\Gamma$ for $z>5$ is currently not well-constrained.
\ifSubfilesClassLoaded{\bibliography{bibliography}{mybib}}
\end{document}

\section{Summary and conclusions}
This paper presents, tests, and illustrates a {\sc python} package called \spec{} to create mock absorption spectra from a hydrodynamical simulation. The code includes routines for writing the input files that {\sc cloudy} \citep{Cloudy23} requires to create interpolation tables for the equilibrium ionisation state of various elements as a function of density and temperature, in the presence of an imposed ionizing background (the \lq ionisation balance\rq\ step). \spec{} then uses {\sc sph} interpolation to compute the ion density, temperature and peculiar velocity for all pixels of a sight line through the simulation volume (the \lq particle projection\rq\ step). Finally, \spec{} computes the optical depth for a range of ions, using the atomic data extracted from the National Institute of Standards and Technology ({\sc nist}) database (the \lq optical depth\rq\ step). \spec{} can include all resonant lines from {\sc nist}, continuum absorption for a range of ions, 
Lyman-series lines for hydrogen and deuterium to almost arbitrary high $n$,
as well as line-broadening from the natural line widths of these lines.
We demonstrate \spec's ability to compute spectra for a sight line traversing the simulation volume once, and combining many such sight lines into one mock observed absorption spectrum, with the user specifying the overall wavelength range and the source's (usually a quasar) redshift. Some features of \spec\ are in common with the {\sc trident} code described by \cite{Hummels17}. Differences include the much larger range of ions used, the column density conserving interpolation scheme, the implementation of higher-order Lyman lines, natural broadening and continuum absorption, and the tight integration with simulation databases.

Mock quasar spectra computed from simulations are crucial for interpreting observed spectra. Example applications include constraining the properties of the intergalactic medium (for example, its thermal state, e.g. \citealt{Schaye00} and ionisation state, e.g. \citealt{Bolton05}), studying the nature of the dark matter (e.g. \citealt{Viel13}), detecting the baryon acoustic oscillation (e.g. \citealt{Debulac15}), or studying gas around galaxies (e.g. \citealt{Turner17}).

We illustrate the versatility of \spec\ with pertinent examples. These include properties of the \lya-forest (mean transmission, flux {\sc pdf} and power spectrum of the transmission). We show that the spectra computed from the \eagle\ simulations \citep{Schaye15} agree very well with data from both high- and low-resolution quasar spectra from $z=0$ to $z=3.5$. We illustrate the different ions that arise when a sight line intersects a damped \lya-absorber ({\sc dla}), and show a full quasar spectrum with the {\sc dla} damping wing and associated continuum absorption. We show that \spec{} reproduces the characteristic feature in the correlation function that arises from the correlated absorption of \ion{H}{i} and \ion{Si}{iii}. Finally, we illustrate the ease with which results from the \eagle\ database \citep{McAlpine16} can be combined with \spec{} to study absorption arising near galaxies.

We use the good agreement of the \lya-forest transmission statistics calculated using \spec{} using the \eagle{} simulation to infer the evolution of the ambient photo-ionisation rate, $\Gamma$. To do so, we first demonstrated that \eagle{} reproduces the evolution of the temperature-density relation. We find that \eagle{} reproduces all these data well over the redshift range $z\in [0, 4]$, when $\Gamma$ is close to the value computed in the model of  \cite{Faucher20} (see Fig.~\ref{fig:Temperature}). At higher $z$, published data on temperature at the mean density, and of the value of $\Gamma$, do not appear to be consistent with each other and more work to understand this is required.

\spec{} uses ionisation fractions that are computed in the optically thin limit. However, the code is also able to use ionisation fractions computed in simulations with radiative transfer ({\sc rt}). We have recently implemented {\sc rt} in the {\sc swift} code \citep{Chan21SPHM1RT}
and coupled this code to the non-equilibrium thermo-chemistry {\sc chimes} network \citep{Richings14}, as described in \cite{Chan26}. We look forward to post-processing these runs with \spec.

\ifSubfilesClassLoaded{\bibliography{bibliography}{mybib}}

\end{document}

\section*{ACKNOWLEDGEMENTS}
This work was supported by the Science and Technology Facilities Council (STFC) through Consolidated Grants ST/P000541/1 and ST/T000244/1 for Astronomy at Durham. 

This work is co-funded by the European Union (Widening Participation, ExGal-Twin, GA 101158446). Views and opinions expressed are however those of the author(s) only and do not necessarily reflect those of the European Union. Neither the European Union nor the granting authority can be held responsible for them. This work used the DiRAC@Durham facility managed by the Institute for Computational Cosmology on behalf of the STFC DiRAC HPC Facility (www.dirac.ac.uk). The equipment was funded by BEIS capital funding via STFC capital grants ST/K00042X/1, ST/P002293/1, ST/R002371/1 and ST/S002502/1, Durham University and STFC operations grant ST/R000832/1. DiRAC is part of the National e-Infrastructure. \spec{} supports outputs of the \textsc{swift} open-source simulation code 
(\url{http://www.swiftsim.com}; \citealt{Schaller24}). We used the {\sc matplotlib} \citep{Hunt07matplotlib}, {\sc numpy} \citep{vand11numpy}, {\sc scipy} \citep{Jone01scipy},
{\sc astropy} \citep{Astropy18}, and {\sc swiftsimio} \citep{Borr20swiftsimio} {\sc python} libraries, and {\sc nasa}'s Astrophysics Data System and {\sc arXiv}
open-access archives.

\section*{DATA AVAILABILITY}
The \eagle\ data used here is generally available, as described in \citep{McAlpine16, Eagle17}.

{\spec{}  is open source. The code, including example Jupyter notebooks, is available at 
\url{https://github.com/specwizard/specwizard}, and the latest stable release can be installed 
from the Python Package Index (\url{https://pypi.org/project/specwizard/}) with 
\texttt{pip install specwizard}.

\bibliographystyle{mn2e}
\bibliography{mn-jour,mybib}

\appendix
\section{Hydrogen photo-ionisation cross section}
\label{sect:sigma}
The fit to the photo-ionisation cross section for \ion{H}{i} is given by \cite{Vern96sigmacross}
as
\begin{align}
    \Sigma(E) &=\Sigma_0\times F(x)\nonumber\\
    F(x)      &= (x-1)^2 \times x^{0.5\,P-5.5} \times \left(1+(\frac{x}{x_a})^{1/2}\right)^{-P}\nonumber\\
    x         &\equiv \frac{E}{E_0}\
    \label{eq:Sigma}
\end{align}
for $E>E_{\rm th}$ where $E$ is the photon's energy and zero otherwise; $E_{\rm th}$ is the binding energy of the ground state. The numerical values of the fit are
\begin{align}
    \Sigma_0 &=  5.475\times 10^{-14}{\rm cm}^2\,;\quad x_a=32.88\nonumber\\
    P &=2.963\,;\quad E_0=0.4298~\unit{eV}\,.
\end{align}

We set $E_{\rm th}=h\,c\,(\mu_e/m_e)\,{\rm Ryd}$, where $h$ is Planck's constant and ${\rm Ryd}$ is the Rydberg constant, 
\begin{align}
    {\rm Ryd}&= \frac{m_e\,e^4}{8\epsilon_0^2\,h^4\,c^2}\approx 1.09737331\times 10^{5}\,\rm {cm}^{-1}\,,    
\end{align}
with $\epsilon_0$ the permittivity of the vacuum; $\mu_e=m_em_p/(m_e+m_p)$ is the reduced mass of the electron in a hydrogen atom in terms of the electron and proton masses, $m_e$ and $m_p$. At the threshold, {\em i.e.} for $E=E_{\rm th}$, the fit yields 
$1/\Sigma(E=E_{\rm thres})=1/ 10^{17.1973}{\rm cm}^2$.

\section{Lyman series line data for Hydrogen \& Deuterium}
\label{sect:hydrogen_data}

Values of the vacuum wavelength, oscillator strength, Einstein $A$-coefficient, and damping constant depend on the properties of the electron wave function. The wave functions are analytical functions when a hydrogenic atom is described with Schr\"odinger's equation, therefore these atomic constants can be computed analytically.

The Rydberg formula for the vacuum wavelength for the transition $n\to n'$ (where $n$ and $n'$ are the principal quantum numbers of the upper and the lower state) is
\begin{align}
    \lambda_{n,n'} &= \left(\frac{1}{n'^2}-\frac{1}{n^2}\right)^{-1}\,R^{-1}\nonumber\\
    R &= \frac{\mu_e}{m_e}\,{\rm Ryd}\,,
    \label{eq:R}
\end{align}
where ${\rm Ryd}$ is the Rydberg constant and $\mu_e$ the reduced mass of the electron, as in the previous Appendix. Spin-orbit coupling breaks the degeneracy of energy levels with respect to the angular quantum number $l$, therefore $\lambda_{nn'}$ additionally 
depends on the values of $l$ and $l'$ of the two states. The wavelength separation of the resulting doublet lines increases rapidly with the atomic number, $Z$, of the nucleus, and is small for hydrogen (and deuterium). We will therefore neglect the $l$-dependence for these elements. Several relativistic effects further affect the \ion{H}{I} wavelengths at a level of $10^{-5}$: this explains the difference between our computed values, shown in Table~\ref{tab:lyman}, and the values from the {\sc nist} database.

Expressions for the oscillator strengths were derived more than a century ago \cite[see e.g.][and original references therein]{Green57, Karzas61},
\begin{align}
    f_{nl, n'l'} &= \frac{1}{3} \frac{{\rm max}(l, l')}{2{\bf l'}+1} \left(\frac{1}{{n'}^2}-\frac{1}{n^2}\right)\,\left(\frac{Z}{a_0} \tau_{nl}^{n'l'}\right)\nonumber\\
    \tau_{nl}^{n'l'} &= \frac{a_0}{Z}\,\left[\frac{\Gamma(n+l+1)\Gamma(n'+l)}{\Gamma(n-l)\Gamma(n'-l+1)}\right]^{1/2}\nonumber\\
    &\times (nn')^{l+1}\,(n+n')^{-n-n'}\,(n-n')^{n-2-l}(n'-n)^{n'-l}\times F\nonumber\\
    F&=\leftindex_{2}{F}_1[l+1-n,l-n';2l,-\frac{4nn'}{(n-n')^2)}]\nonumber\\
    &-\left(\frac{n-n'}{n+n'}\right)^2\,
    \leftindex_{2}{F}_1[l-1-n, l-n'; 2l, -\frac{4nn'}{(n-n')^2}]\,,
\end{align}

for transitions that satisfy $l'-l=\pm 1$. These are equations 35 and 41 in \citet{Karzas61}; the atomic number $Z$ and Bohr radius $a_0$ cancel out in the calculation of the oscillator strength. We have written ${\bf l'}$ in the first line in bold face: this factor is $l$ in \cite{Karzas61} but $l'$ in \cite{Green57}: we suspect that the first value is a typo. In these expressions, $\Gamma$ is the gamma-function and $\leftindex_{2}{F}_1$ is the hypergeometric function\footnote{Implemented as \texttt{hyp2f1} in {\sc Python}'s {\sc scipy} library.}. A naive implementation of these expressions results in overflows when $n\gtrapprox 150$. These are easily resolved by applying Stirling's approximation in the calculation of $\Gamma$ for high $n$, and rewriting the product of powers of $n$ and $n'$ in terms of powers of $n$ and powers of $n'/n$. 

In the particular case of hydrogen or deuterium Lyman lines, the expressions in {\sc si} units for the oscillator strength, $f_{n,1}$, and Einstein-A coefficient, $A_{n,1}$, for transitions between $1s$ and $np$  simplify to
\begin{align}
    f_n & \equiv f_{n,1} = \frac{256}{3}\,\frac{1}{n^3}\,\left(1-\frac{1}{n}\right)^{2n-4}\,\left(1+\frac{1}{n}\right)^{-2n-4}\nonumber\\
    A_n &\equiv A_{n,1} = \frac{\pi e^{10} \mu_e}{96\,c^3\,\epsilon_0^5 h^6}\,f_{n,1}\,\left(1-\frac{1}{n^2}\right)^2\,,
    \label{eq:analytical}
\end{align}
with $f_{n,1}\equiv f_{n1, 10}=f_{np, 1s}$. For deuterium, $m_p\to 2.014~\unit{amu}$ in the expression for $\mu_e$. These are equations 9 and 10 in \cite{Mortlock16} (see that paper for original references), except we replaced $m_e$ by $\mu_e$. With this substitution, values of $f_n$, $\lambda_{n,1}$ and $A_n$ agree to better than 0.01~per cent with those quoted in the {\sc nist} database. However, with these equations in hand, we can compute them for arbitrarily large values of $n$. To compute the damping width from Eq.~(\ref{eq:gamma}) requires Einstein coefficients for different values of $l$ and $l'$: we use the code described by \cite{hoang-binh_program_2005} to compute these. We find better than sub-percent agreement in the resulting Einstein values for transitions from levels $n1$ to $1s$ for which Eq.~(\ref{eq:analytical}) applies.

\label{lastpage}
\bsp
\end{document}